%% file: cl01.2.tex
\documentclass{aastex}
\usepackage{natbib,epsfig,emulateapj5,rotate}

\tightenlines

\begin{document}
\newcommand{\be}{\begin{eqnarray}}
\newcommand{\ee}{\end{eqnarray}}
\newcommand{\hii}{H{\sc ii }}
\newcommand{\hi}{H{\sc i }}
\newcommand{\etal}{et al.}
\def\llbdm{{\cal L}_{\rm \ell,b,\DM}}
\def\lpsrd{{\cal L}_{\rm d}}
\def\lpsrs{{\cal L}_{\rm psr,s}}
\def\lgals{{\cal L}_{\rm gal,s}}
\def\lxgals{{\cal L}_{\rm xgal,s}}
\def\like{{\cal L}}
\def\ls{{\cal L}_s}
\newcommand{\lampsrd}{{\Lambda}_{\rm psr,d}}
\newcommand{\lampsrs}{{\Lambda}_{\rm psr,s}}
\newcommand{\lamgals}{{\Lambda}_{\rm gal,s}}
\newcommand{\lamxgals}{{\Lambda}_{\rm xgal,s}}
\def\lam{{\Lambda}}
\def\dl{{\rm D_L}}
\def\du{{\rm D_U}}
\newcommand{\Deff}{D_{\rm eff}}
\def\norm{{\cal N}}
\def\npsrd{{N_{\rm psr,d}}}
\def\dhat{{\hat D}}
\def\DM{{\rm DM}}
\def\RM{{\rm RM}}
\def\SM{{\rm SM}}
\def\EM{{\rm EM}}
\def\DMinfty{{\rm DM}_{\infty}}

\def\SMxgal{{\rm SM}_{\theta,x}}
\def\SMgal{{\rm SM}_{\theta,g}}
\def\SMtau{{\rm SM}_{\tau}}

\def\smun{{kpc \,\, m^{-20/3}}} 
\def\cnsq{{C_n^2}} 
\def\nbar{{\overline{n}_e}}
\def\dne{{\delta n_e}}
\def\pne{{P_{\delta \rm n_e}}}
\newcommand{\dphi}{D_{\phi}}

\def\dnud{{\Delta\nu_{\rm d}}}
\def\taud{{\tau_d}}
\def\narms{{N_{\rm arms}}}
\def\nclumps{{N_{\rm clumps}}}
\def\nvoids{{N_{\rm voids}}}

\def\nparms{{N_{\rm parms}}}
\def\nmc{{N_{\rm MC}}}

\def\xvec{{\bf x}}
\newcommand{\xbar}{\overline {x}}
\newcommand{\ybar}{\overline {y}}
\newcommand{\zbar}{\overline {z}}

\newcommand{\rc}{r_{\rm c}}
\newcommand{\dc}{d_{\rm c}}
\newcommand{\nec}{n_{\rm e_c}}
\newcommand{\Fc}{F_{\rm c}}

\newcommand{\sgra}{\mbox{Sgr~A${}^*$}}
\newcommand{\dgc}{\ensuremath{D_{\mathrm{GC}}}}
\newcommand{\delgc}{\ensuremath{\Delta_{\mathrm{GC}}}}
\newcommand{\rsun}{\ensuremath{\mathrm{R}_\odot}}
\newcommand{\wlism}{\ensuremath{w_{\rm lism}}}
\newcommand{\wlhb}{\ensuremath{w_{\rm lhb}}}
\newcommand{\wlsb}{\ensuremath{w_{\rm lsb}}}
\newcommand{\wldr}{\ensuremath{w_{\rm ldr}}}
\newcommand{\wloopI}{\ensuremath{w_{\rm loop I}}}
\newcommand{\wvoids}{\ensuremath{w_{\rm voids}}}
\newcommand{\halpha}{\ensuremath{\rm H\alpha}}
\newcommand{\nelism}{n_{\rm lism}}
\newcommand{\neldr}{n_{\rm ldr}}
\newcommand{\nelsb}{n_{\rm lsb}}
\newcommand{\nelhb}{n_{\rm lhb}}
\newcommand{\neloopI}{n_{\rm loop I}}
\newcommand{\nevoids}{n_{\rm voids}}
\newcommand{\neclumps}{n_{\rm clumps}}
\newcommand{\Flhb}{F_{\rm lhb}}
\newcommand{\Fldr}{F_{\rm ldr}}
\newcommand{\Flsb}{F_{\rm lsb}}
\newcommand{\FloopI}{F_{\rm loop I}}
\newcommand{\Flism}{F_{\rm lism}}
\newcommand{\negal}{n_{\rm gal}}
\newcommand{\negc}{n_{\rm gc}}
\newcommand{\nearms}{n_{\rm arms}}
\newcommand{\ok}{$\surd$}
\newcommand{\bvec}{{\bf b}}
\newcommand{\nutrans}{\nu_{\rm trans}}
\newcommand{\nover}{N_{\rm over}}
\newcommand{\nhits}{N_{\rm hits}}
\newcommand{\nlum}{N_{\rm lum}}
\newcommand{\smin}{S_{\rm min}}
\newcommand{\Dmax}{D_{\rm max}}
\newcommand{\Lp}{L_{\rm p}}
\newcommand{\lp}{l_{\rm p}}
\newcommand{\fLp}{f_{\Lp}}
\newcommand{\flp}{f_{\lp}}
\newcommand{\fLpP}{f_{\lp, P}}
\newcommand{\fLpDM}{f_{\Lp, \DM}}
\newcommand{\flpDM}{f_{\lp, \DM}}
\newcommand{\fDM}{f_{\DM}}
\newcommand{\fD}{f_{\rm D}}
\newcommand{\RGC}{R_{\mathrm{GC}}}
\newcommand{\HGC}{H_{\mathrm{GC}}}
\def\data{{\cal D}}
\def\models{{\cal M}}
\def\info{{\cal I}}
\newcommand{\thetavec}{{ \mbox{\boldmath $\theta$} }}
\newcommand{\rperp}{r_{\perp}}
\newcommand{\cu}{C_u}
\newcommand{\csm}{C_{\rm SM}}
\newcommand{\rscatt}{r_{\rm scatt}}
\newcommand{\nebar}{\bar n_e}
\def\diffus{{\cal D}} 
\def\vartheta{V_{\theta}} 
\def\cnsq{C_n^2} 
\def\FWHM{{\theta_{d}}} 
\newcommand{\lmax}{\ell_{\rm max}}
\newcommand{\bmax}{b_{\rm max}}


\title{ 
NE2001. II. 
Using Radio Propagation Data to Construct a Model for 
the Galactic Distribution of Free Electrons}
\author{J. M. Cordes}
\affil{Astronomy Department and NAIC, Cornell University,
 Ithaca, NY~~14853\\ cordes@spacenet.tn.cornell.edu}
\bigskip
\author{T.~Joseph~W.~Lazio}
\affil{Naval Research Lab, Code~7213, Washington, D.C. 20375-5351  \\
Joseph.Lazio@nrl.navy.mil}

\bigskip
\begin{abstract}
In Paper I we present a new model for the Galactic distribution of free
electrons. In this paper we describe the input data and methodology for
determining the structure and parameters of the model.  
We identify lines of sight on which  discrete regions either enhance
or diminish the dispersion or the scattering.   
Most do not coincide with known 
\ion{H}{2} regions or supershells, 
most likely because the enhancements correspond to 
column densities smaller than detection thresholds for the emission measure
in recombination-line
surveys.   
\end{abstract}

\section{Introduction}\label{sec:intro}

In Paper I (Cordes \& Lazio 2002) we present a new model for the Galactic
electron density and its fluctuations.   This model aims to quantify
the density of thermal electrons in the Galaxy which may be used 
for inverting pulsar dispersion measures into pulsar distances.   It
also quantifies small-scale electron density fluctuations
that are responsible for the scattering and scintillations of
compact radio sources, ranging from Galactic pulsars and masers,
and Sgr A* to active galactic nuclei (AGN)  and gamma-ray burst
(GRB) afterglows in other galaxies.

In this paper,  we present the data and the formalism used to construct the
model.  We describe the formalism that relates measureable quantities
to line of sight measures that provide the basis for testing models.
We also describe
methods used to test the model, including a likelihood analysis,
and we demonstrate the need for inclusion of various
model components using the likelihood function and other measures
for the goodness of fit.  Finally, we describe lines of sight
that provide especially important constraints on the model 
or that otherwise require special treatment through inclusion of
clumps or voids somewhere along the propagation path. 

Our model is constructed for multiple purposes.  First and foremost
is to allow estimation of pulsar distances and scattering.  In this
regard, the number of parameters of the model is unrestricted by
any notions of parsimony:  good estimates are the bottom line.  
However, a conflicting goal --- determining
the ionized structure of the Galaxy  on a variety of length scales--- does    
require considerations of model complexity and whether specific features are
demanded by the data.    

The organization of the paper is as follows.  Observables and integrated
measures are discussed in \S\ref{sec:obs}.   
In \S\ref{sec:distribution} we discuss the components of the model
that are required by the data. 
We use straight-forward empirical distributions
to argue for the existence of the various components. 
\S\ref{sec:model} presents the detailed model.
\S\ref{sec:methodology} discusses our fitting methods and figures
of merit.
\S\ref{sec:comparison} compares results of model fitting when we
exclude one or more components from the model.  We also compare
NE2001 with TC93 and a more recent model of 
G{\'o}mez, Benjamin, \& Cox (2001; hereafter GBC01).
\S\ref{sec:goodness} discusses additional aspects of model
performance.
\S\ref{sec:LOS} discusses particular lines of sight that are
either problematic or especially informative.
\S\ref{sec:model.discussion} presents limitations of the model
and 
\S\ref{sec:discussion} summarizes our results and future 
models that we anticipate.
In Appendix~\ref{app:sm} we derive estimators for the scattering
measure from relevant scintillation and scattering observables.
Tables in the Appendix give scattering data for the lines of sight
in our database of measurements.

\section{ Observable Quantities and Integrated Measures}
	\label{sec:obs}

In Paper~I we summarized the kinds and number of input data used for our model.
They include measurements on pulsars of
the dispersion measure, \DM, and various scattering observables of
pulsars and other compact Galactic and extragalactic sources
that include the angular broadening, $\theta_d$, the pulse broadening
time $\tau_d$ and the scintillation bandwidth, $\dnud$.   In our analysis,
we consider scattering observables only in the so-called strong-scattering
regime (Rickett 1990; Cordes \& Lazio 1991) 
where wavefronts from
sources with a high degree of spatial coherence are perturbed by
much more than 1 rad (rms) when propagating through the ionized 
interstellar medium (ISM).  In particular, the phase 
{\it difference} between two ray paths separated by the Fresnel scale 
($\sim \sqrt{\lambda D} \approx 10^{11}$ cm), where 
$\lambda$ is the electromagnetic wavelength and $D$ is a characteristic
distance through the medium,  
is much larger than 1 rad in order that scattering effects occur. 
In strong scattering, there are no unscattered components to the source's
wavefield and the diffracted intensity is correlated over a
frequency scale (the `scintillation bandwidth') $\dnud \ll \nu$ where
$\nu=c/\lambda$.   In Appendix~\ref{app:sm}, we discuss how the scattering
observables may be used to estimate the scattering measure,
\SM, in the strong-scattering regime.

For completeness, we point out that there are other scattering
observables, but they are of less utility.  The characteristic
scintillation time scale~$\Delta t_d$ is not particularly useful for
quantifying \SM\ because an unknown characteristic velocity is also
needed to make the relation between~$\Delta t_d$ and~\SM.  No
spectral-line Galactic or extragalactic sources whose emission line
widths are narrow enough to display spectral broadening are known,
though future detection of extraterrestrial transmitters may change
this situation \cite{cl91}.  Finally, measurements exist for some
lines of sight of scintillations in the weak-scattering regime which
also can yield estimates for \SM.  Rickett, Coles, \& Markkanen (2000)
describe such measurements and how to calculate \SM\ in that regime.
 
Independent distance measurements of Galactic sources, particularly
pulsars but also masers, microquasars and Sgr~A*, are also crucial
for calibrating the model.   As described in Paper I, we cast these
in the form of a distance range, $[\dl, \du]$. 

In Figure 1 and Table 1 of Paper I, we summarize the numbers of
available measurements, which include upper bounds as well as interval
constraints.  In Tables~\ref{tab:psr.smtable}--\ref{tab:xgal.smtable} we tabulate
scattering measures and relevant input data for lines of sight toward
pulsars, other Galactic sources, and extragalactic sources.

\section{Basic Structure of the Model Based on Empirical Distributions}
	\label{sec:distribution}

The electron density model presented in Paper I contains a number
of components that are suggested by the data.    These are:
(a) a thick disk with large Galactocentric scale height;
(b) a thin, annular disk in the inner Galaxy;
(c) spiral arms;
(d) a Galactic center component;
(e) structures in the vicinity of the Sun that we collectively refer to
as the local ISM component;
and 
(f) other over-or-underdense regions that we refer to as clumps and voids. 
These salient features  may be identified
through investigation of observable quantities plotted against 
Galactic coordinates and other variables, as we discuss in this section. 
Detailed fitting of the parameters of the components and identification
of lines of sight that intersect voids and clumps are discussed in later
sections.

\subsection{\DM\ and Independent Distance Constraints}

From {\DM}s and independent distance constraints, we may infer the
scale height of the thick disk close to the Sun, the need for a thin
disk and spiral arms, and the existence of enhancements and deficits
of electron density near the Sun.


\subsubsection{Evidence for a Thick Disk}\label{sec:thick_evidence}

Figure~\ref{fig:dmz} shows $\DM\sin\vert b\vert$ plotted against
$\vert z\vert$ for pulsars with independent distant estimates; similar 
versions of this figure, though based on fewer data, have been
discussed by Reynolds~(1991); Nordgren, Cordes \& Terzian (1992); and GBC01.
Schematic, plane-parallel  models for the electron density are overplotted
for different mid-plane densities and scale heights.  Figure~\ref{fig:dmz}
demonstrates that 
(a) the average electron density along a given line of sight
is typically in the range of~0.01--0.1~cm$^{-3}$, as indicated
by the dashed lines;
(b) the paucity of Galactic pulsars with  
$\DM\sin\vert b\vert \gtrsim 25$ pc cm$^{-3}$ indicates that
the free electron distribution is bounded in the $z$ direction
with a scale height $\sim 1$ kpc;
and 
(c) the variations  in mean density are larger than can be explained
by distance errors, signifying changes in
electron density on a variety of scales.
For a plane parallel medium, examples of which are shown as
solid lines in Figure~\ref{fig:dmz},  the midplane density is
about~0.02--0.04~cm$^{-3}$ with (exponential) scale height
$\sim 0.75\pm 0.25$ kpc. 
Our model described below is considerably more complex than
a plane-parallel medium so these estimates are only
by way of example.
These results echo the statements of Kulkarni \& Heiles~(1988) that
simple plane parallel models are ``only statistical in nature and
should not be interpreted too literally.''

\smallskip
\epsfxsize=8truecm
\epsfbox{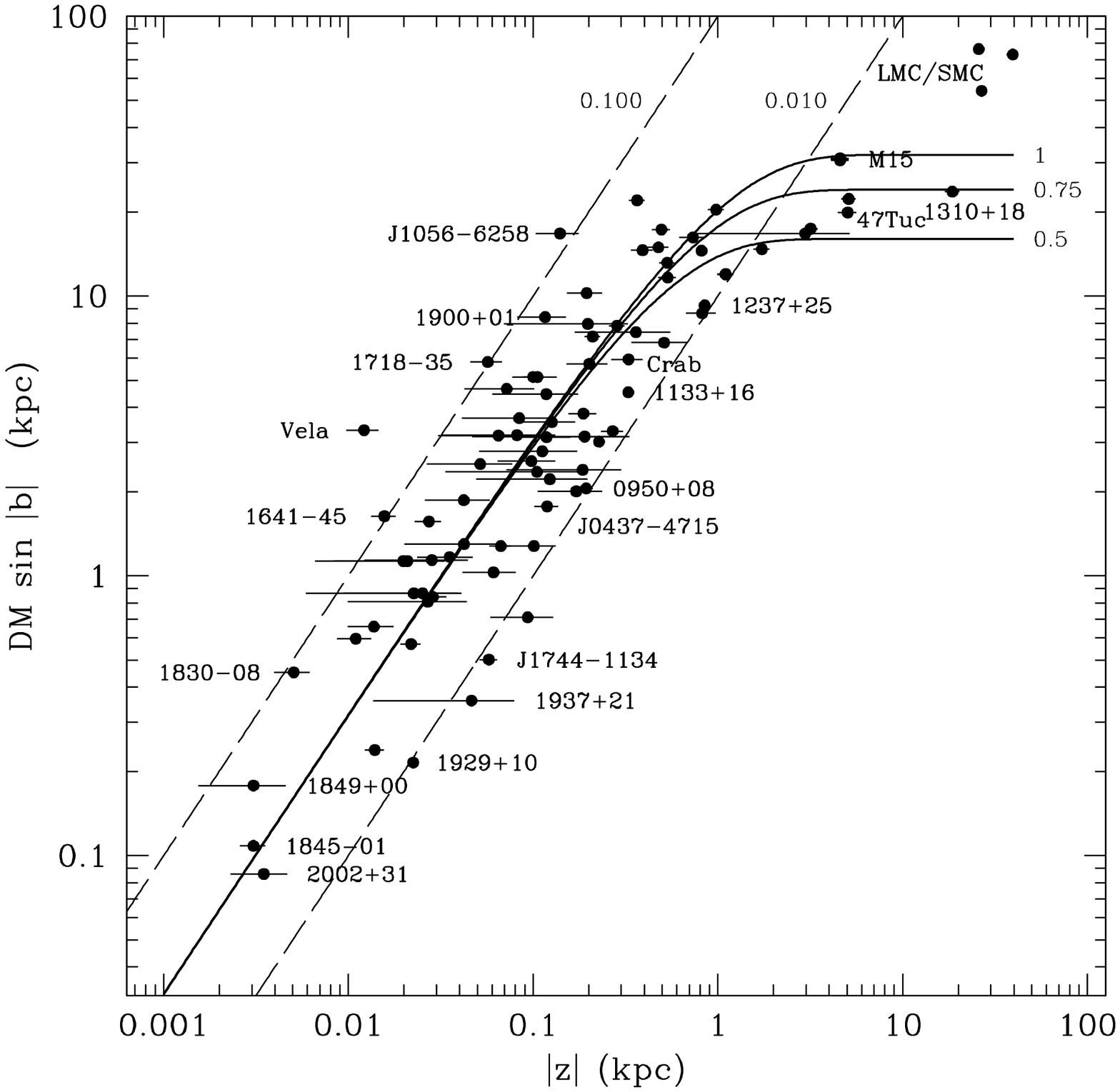}
\figcaption{
\label{fig:dmz}
Component of \DM\ perpendicular to the Galactic plane,
$\DM\sin \vert b \vert $, plotted against $z$ distance for those pulsars
having independent distance constraints of the form
$[\dl, \du]$ and where the implied $z$ range spans less
than one decade.   The dashed lines show curves for
media with constant $n_e = 0.01$ and 0.1 cm$^{-3}$.
The solid lines represent plane parallel models with
exponential scale heights of 0.5, 0.75 and 1 kpc
(as labelled).  For these models, the midplane density
is held fixed at 0.032 cm$^{-3}$.
The paucity of objects with 
$\DM\sin \vert b \vert \gtrsim 25$ pc cm$^{-3}$,
apart from those in the LMC and SMC with resultant
contamination of their DMs,  is consistent with
a well defined truncation of the Galaxy's free electron
layer at or near  an effective scale height of 1 kpc.
}
\medskip

\subsubsection{The Thin, Inner-Galaxy Disk and Spiral Arms}

The thin disk and spiral arms can be inferred from the distribution
of \DM\ in Galactic longitude~$\ell$.
Figure~\ref{fig:lbplot8} shows \DM\, plotted against $\ell$ for
8 ranges of $b$.   The mean \DM\ in each frame  falls off 
monotonically with mean $b$, indicating the existence of a 
disk-like distribution.  
The structure evident in Figure~\ref{fig:lbplot8} reflects
(a)~the existence of a small scale-height region in the inner Galaxy
that is responsible for large DMs at low Galactic latitudes,
and 
(b)~a scale height for the inner disk component $\sim 0.15$ kpc. 
At high latitudes, there is little systematic variation with $\ell$.
At low latitudes, the largest values of \DM\, are in
the longitude range $-90\arcdeg \lesssim \ell \lesssim 70\arcdeg$.
The asymmetry of this range about $\ell = 0\arcdeg$ has been noted by others 
(Johnston \etal\, 1992; TC93) and was explicitly used by TC93 
to aid in the definition and fitting of spiral arms. 
With more than twice the number of pulsars now available, the asymmetry
is even more striking and again suggests the presence of spiral
structure as the most reasonable cause for the asymmetry.  Selection
effects are unlikely to be the cause of the asymmetry as a number of
deep searches at low latitudes near $\ell \approx 60\arcdeg$ have been
carried out with the Arecibo Observatory, yet the number of pulsars
around $\ell \approx -60\arcdeg$ is clearly much higher.  
An alternative suggestion is that a Galactic bar (e.g., Blitz \&
Spergel 1991) underlies this asymmetry.  However, the typical size and
orientation usually attributed to the bar suggest that its effects
would appear for longitudes in the range $\sim \pm 20\arcdeg$, much
smaller than the range in which asymmetry is seen.

\smallskip
\epsfxsize=8truecm
\epsfbox{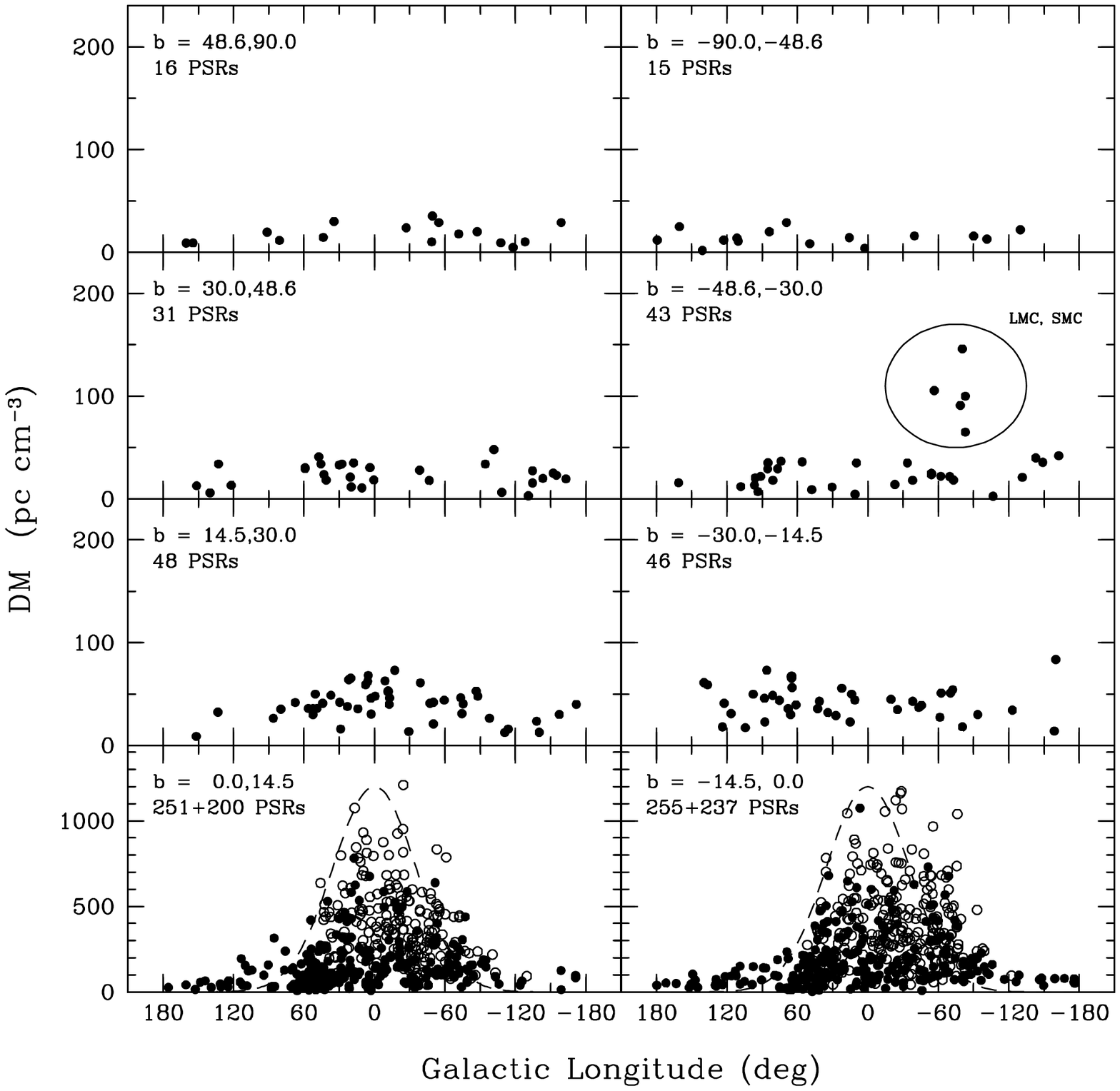}
\figcaption{
\label{fig:lbplot8}
Scatter plots of \DM\, vs. Galactic longitude for eight ranges of
latitude.  Filled circles: data from the Princeton pulsar
catalog (Taylor, Lyne, \& Manchester 1993 with updates);
open circles: data from the Parkes Multibeam survey.
The latter covers only low latitudes and thus
contributes only to the bottom two panels.  
Note that the \DM\, scale for the bottom two panels
is different from the others.
The circle in the right, second-from-top panel
indicates pulsars whose DMs receive contributions from 
the LMC and SMC. 
The dashed lines in the bottom two panels show a Gaussian
function with $1/e$ width of~{50\arcdeg}  centered on $\ell = 0\arcdeg$.
It helps delineate the asymmetry of the 
data points about $\ell = 0\arcdeg$.
}
\medskip

Figure~\ref{fig:blplot8} shows the latitude dependence of \DM\ for
eight longitude ranges.  Here, again, the strong longitude dependence
at low latitudes is evident.  The upper envelopes of \DM\ values
appear to be well described, though not perfectly, by a plane-parallel
model that yields $\DM_{90}\vert \csc(b)\vert$ with $\DM_{90}\approx
24$ pc cm$^{-3}$ (apart from pulsars in the LMC and SMC).  As
described below, this value for $\DM_{90}$ is consistent for many
lines of sight with our formal model fitting, though our model is not
plane-parallel in form.

\medskip
\epsfxsize=8truecm
\epsfbox{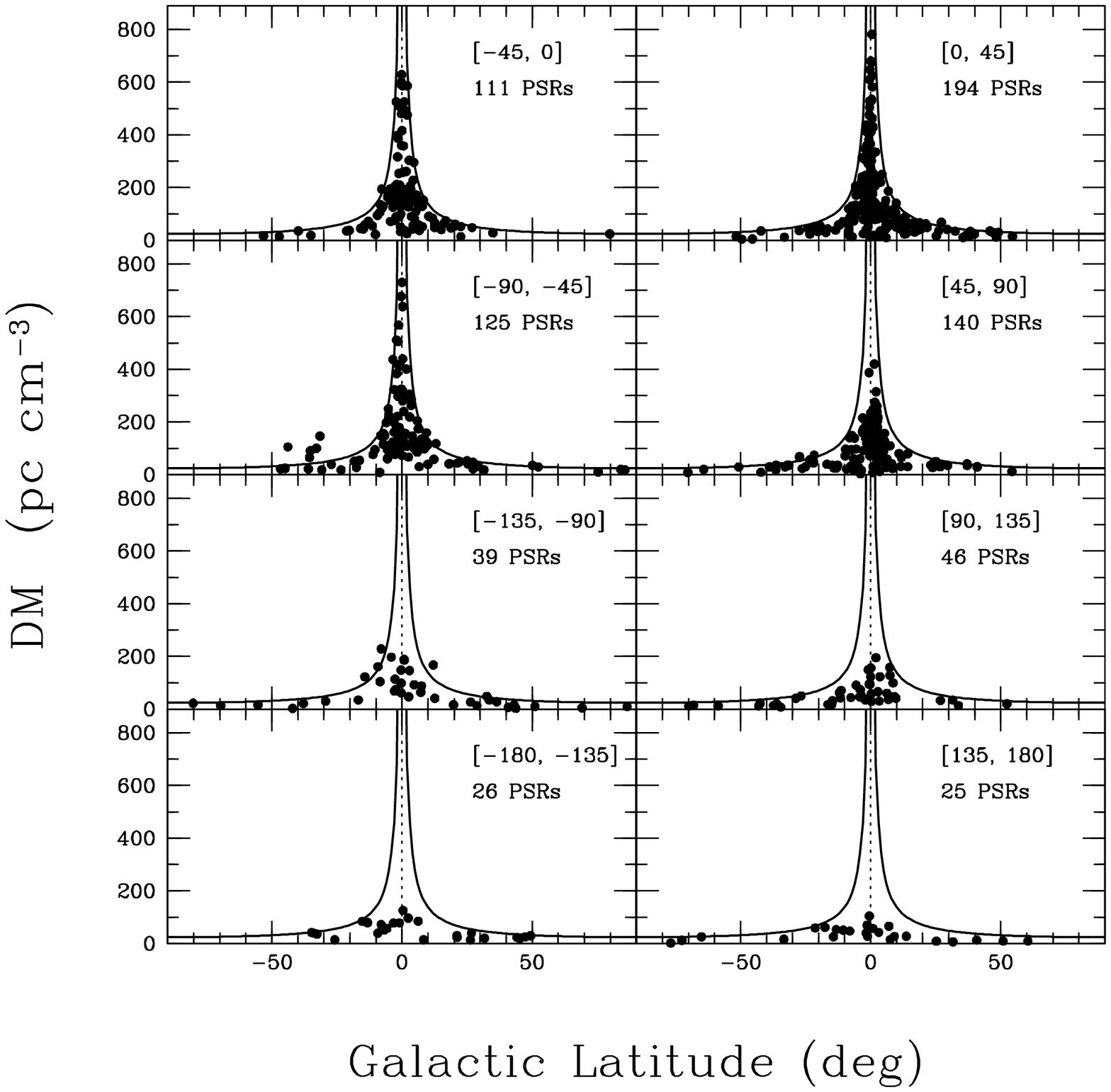}
\figcaption{
\label{fig:blplot8}
Scatter plots of \DM\, vs. Galactic latitude for ranges of
longitude.  The data are from the Princeton catalog.
The solid lines represent the maximum \DM\, for a plane
parallel medium with $\DM = 24$ pc cm$^{-3}$ for
$\vert b \vert = 90\arcdeg$.  
}
\medskip

Taken together, Figures~\ref{fig:lbplot8}--\ref{fig:blplot8} require
the existence of an inner Galaxy region with electron density larger
than and scale height smaller than those of the thick disk.
Physically, this region could be made up solely of spiral arms, of an
annular disk like the molecular ring, or a combination of the two, as
was chosen in TC93 and which we follow here.

Figure~\ref{fig:MW} shows  schematically 
an inner-Galaxy disk and how it is expected to be manifested for lines of sight
with longitudes smaller than $\lmax \approx \sin^{-1} r_d /\rsun$  
and latitudes smaller than $\bmax \approx \sin^{-1} H_d / (\rsun-r_d)$,
where $r_d$ and $H_d$ are the characteristic radius and scale
height of the disk.
Evaluating using $\lmax \approx 60\arcdeg$ and $\bmax \approx 2\arcdeg$,
we find that $r_d \approx 4.3$ kpc and $H_d \approx 0.15$ kpc.
These estimates for the radius and scale height of an inner-Galaxy
disk correspond to our formal model parameters, $H_2$ and $A_2$. 

\medskip
\epsfxsize=8truecm
\epsfbox{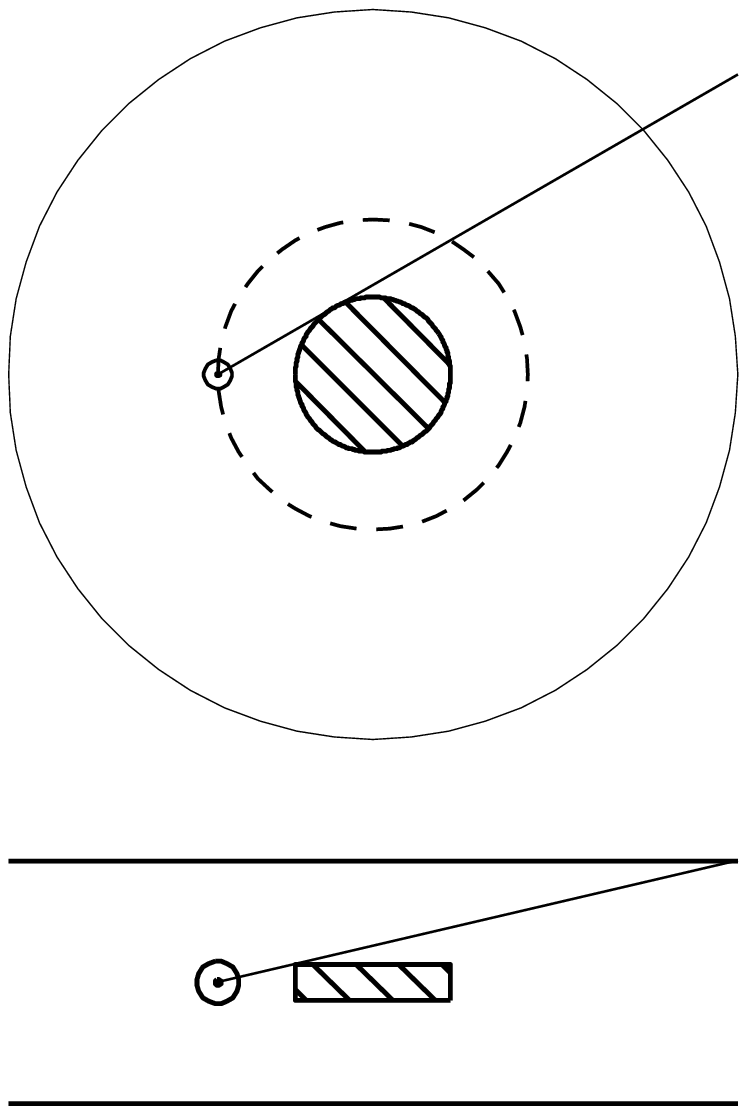}
\figcaption{
\label{fig:MW}
Schematic views of the Galaxy.
Top: view from above the Galactic plane showing the thin, inner-Galaxy
component (hatched) and thick outer Galaxy disk.
A line of sight is shown emanating from the solar circle with
radius $\rsun$ (dashed circle).
Bottom: view of the Galaxy from the side.
}
\medskip

While we have argued for the presence of spiral structure in $n_e$
using the asymmetry of \DM\ in Galactic longitude, a stronger argument
can be made for spiral structure by investigating residual dispersion
measures, as we have discussed in \hbox{Paper~I}.  Briefly, one can compare
the difference between the \DM\ integrated to infinite distance in a
model and the pulsar \DM.  Doing so for axisymmetric models (Figure~4, 
Paper~I) yields clearly insufficient electron column densities along
lines of sight populated by other spiral arm tracers (e.g., \ion{H}{2} 
regions, radio recombination lines).  We take these deficiences from
an axisymmetric model to be an indication that spiral structure is a
necessary component of any complete Galactic $n_e$ model.

\subsubsection{Evidence for Strong Fluctuations}

Figure~\ref{fig:nebar} shows $\nebar \equiv \DM/D$, where
$D$ is constrained to an interval $[\dl, \du]$ from measurements 
independent of \DM.    It is obvious that $\nebar$ varies significantly
between lines of sight, by more than two orders of magnitude.
Some \DM\ excesses are seen on 
short paths, such as the line of sight to 
the Vela pulsar $\sim 0.25$ kpc away, while others are significant fractions of the
distance to the Galactic center (assumed to be 8.5 kpc).
The estimates of $\nebar$ therefore show evidence for strong departures
from a local electron density that varies only smoothly with location.
Such results are not surprising given the known complexity of the
phase structure of the ISM, but they imply that a model for the electron
density must contain small scale structure combined with large-scale
structure.

\medskip
\epsfxsize=8truecm
\epsfbox{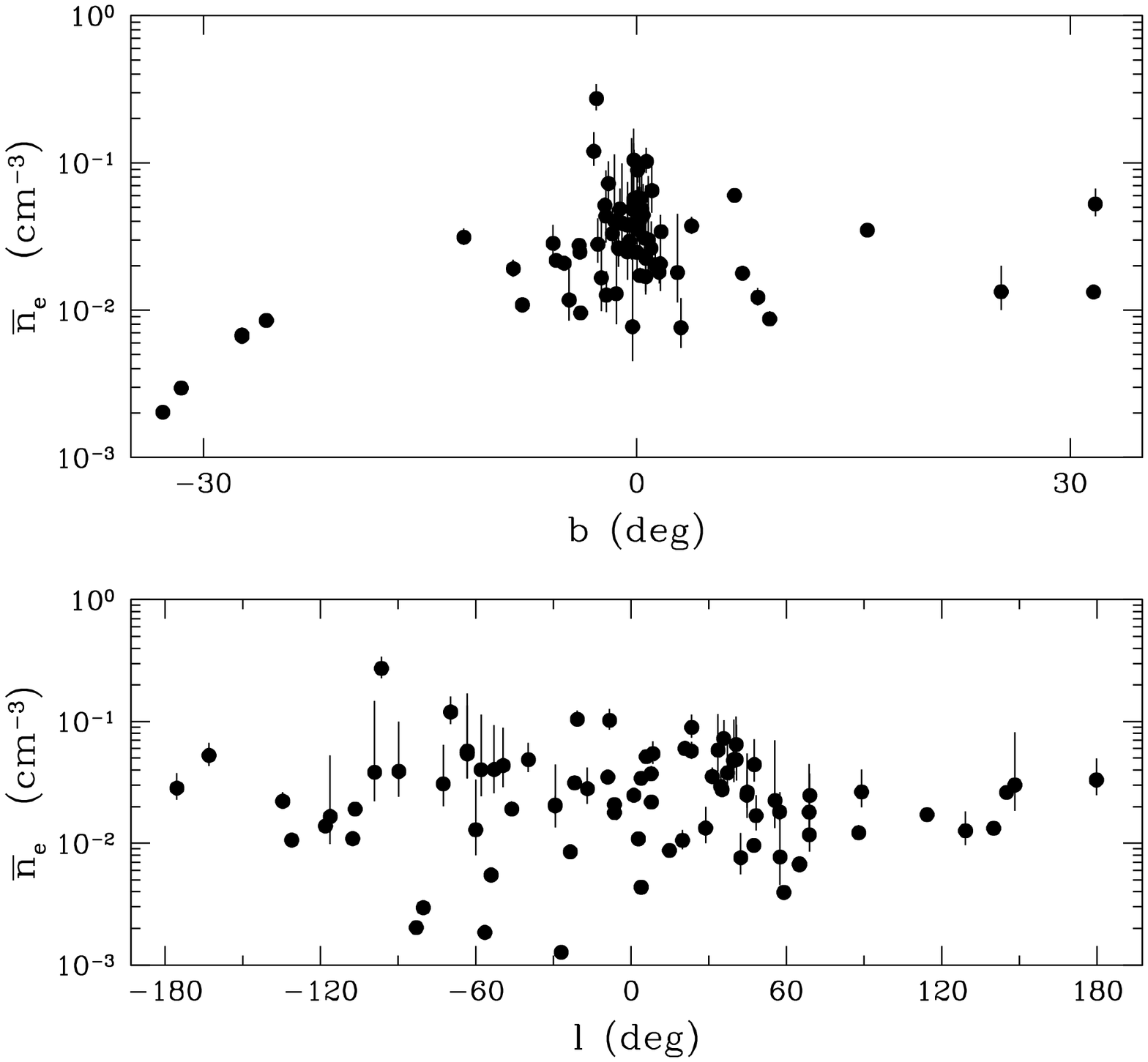}
\figcaption{
\label{fig:nebar}
Plot of $\nebar$ against Galactic latitude and longitude for pulsars
with independent distance contraints, $[\dl, \du]$. The points designate
\DM\ divided by the mean distance while the plotted bars, some of which are
too small to be visible, designate the range of allowed $\nebar$.
The smallest values are for pulsars in the Magellanic clouds and in 
globular clusters, the lines of sight to which largely  intersect regions  
outside the Galactic disk with low electron density.
}
\medskip

\subsection{Constraints from Scattering Measurements}

The same electrons responsible for causing the dispersion of pulsar
signals also appear to be responsible for the observed scattering
\cite{cwf+91}.  While scattering measurements are fewer in number
than the sample of DMs, they provide additional lines of sight to help
constrain the model and cover the sky sufficiently well that the
scattering properties of the ISM can be modelled also.

Figure~\ref{fig:smvsb} shows that the scattering measure has a
distribution in~$b$ that is only approximately described by a cosecant law.
The solid line is a cosecant law 
$\SM(b) = \SM_{90}\vert \csc b\vert$
for an exponential $z$ falloff, where we relate \SM\ to \DM\
using the formalism of Paper~I (equations~[11]--[13]):
\be
\SM_{90} = \csm~ F\, \DM_{90}^2 / 2H
	\approx 10^{-6.04} F\, \DM_{90}^2/H,
\label{eq:smcsc}
\ee
where the second equality holds for \DM\ and \SM\ in standard units.
Using
$\DM_{90} \approx 24$ pc cm$^{-3}$, a scale
height of 0.95 kpc and  a fluctuation parameter $F=0.1$ we obtain 
$\SM_{90}\approx 10^{-4.25}$ kpc m$^{-20/3}$ at $\vert b \vert = 90\arcdeg$. 
This asymptotic value for $\SM_{90}$ is about right, but there are strong departures 
from this model both
at low and high latitudes.  Evidently the ISM is patchy everywhere.
At low latitudes, values of \SM\ are much larger than 
the model, indicating the clear need for 
a thin-disk component for scattering material, 
in accord with
previous studies indicating that the most  intense scattering is associated with
extreme Population I activity in the Galaxy 
(e.g., Cordes, Weisberg, \& Boriakoff~1985).
At high latitudes, $\vert b\vert > 15\arcdeg$, 
\SM\ shows little systematic variation with  latitude, 
suggesting the presence of a thick disk component of scattering material.

\medskip
\epsfxsize=8truecm
\epsfbox{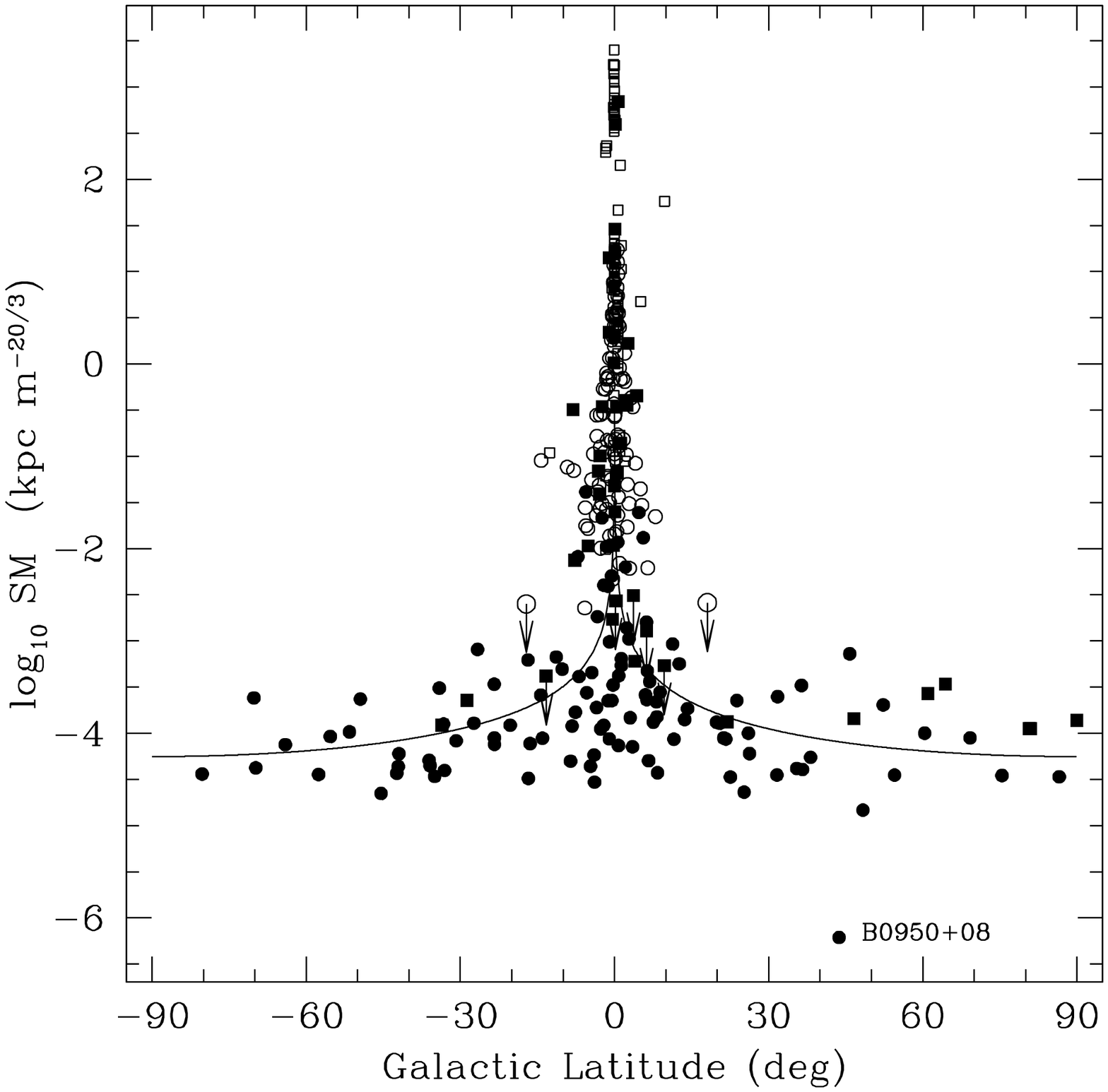}
\figcaption{
\label{fig:smvsb}
Scatter plot of $\log$ \SM\, vs. Galactic latitude.
Values for \SM\ depend in some cases on a distance estimate
which is obtained from the model presented in this paper.  
Open circles: pulse broadening times ($\taud)$;
Filled circles: scintillation bandwidths ($\dnud$);
Open squares: Scattering diameters ($\theta_d$) of Galactic sources;
Filled squares: Scattering diameters of extragalactic sources.  
The solid line is a cosecant model for a plane parallel medium
with the same $\DM_{90} = 24$ pc cm$^{-3}$  
as in Figure \ref{fig:lbplot8},
a scale height of 0.95 kpc (for $n_e$), and a fluctuation
parameter $F=0.1$, combined according to Eq.~\ref{eq:smcsc}.
This plot shows measured values and only a few  upper bounds; 
nonconstraining upper bounds are excluded for clarity.
}
\medskip

One might question whether the high latitude measurements signify
the existence of a very large Galactic halo of scattering material
or intergalactic scattering.  The answer is that high latitude 
scattering is dominated by a thick ($\sim 1$ kpc) disk because the
scattering measures of pulsars above the disk component
are nearly identical to the values of SM for extragalactic sources.
This conclusion is consistent with the argument we gave in the
discussion of Figure~\ref{fig:dmz}.

Investigation of pulse broadening as a function of \DM\
yields nearly direct evidence for strong Galactic structure
in the scattering medium.
Pulse broadening is, of  course,  correlated with \DM, but not perfectly.
Figure \ref{fig:plottau}
shows $\taud$ plotted against DM.
We have scaled all available measurements to 1 GHz using a frequency
dependence $\tau_d \propto \nu^{-4.4}$, the scaling law for a
Kolmogorov medium with negligible inner scale (see Appendix).
Also shown is  a
parabolic fit (solid line) and $\pm 1.5\sigma$ lines (dashed).
The fit is with $\log DM$ as the independent variable:
\be
\log \taud (\mu s) = -3.59 &+& 0.129\log \DM + 1.02 (\log \DM)^2
	\nonumber \\
   &-& 4.4\log\nu_{\rm GHz}.
\label{eq:taudvsdm}
\ee
The scattering time for individual pulsars
can deviate considerably  from the fit,  with $\sigma_{\log\tau} = 0.65$.

\medskip
\epsfxsize=8truecm
\epsfbox{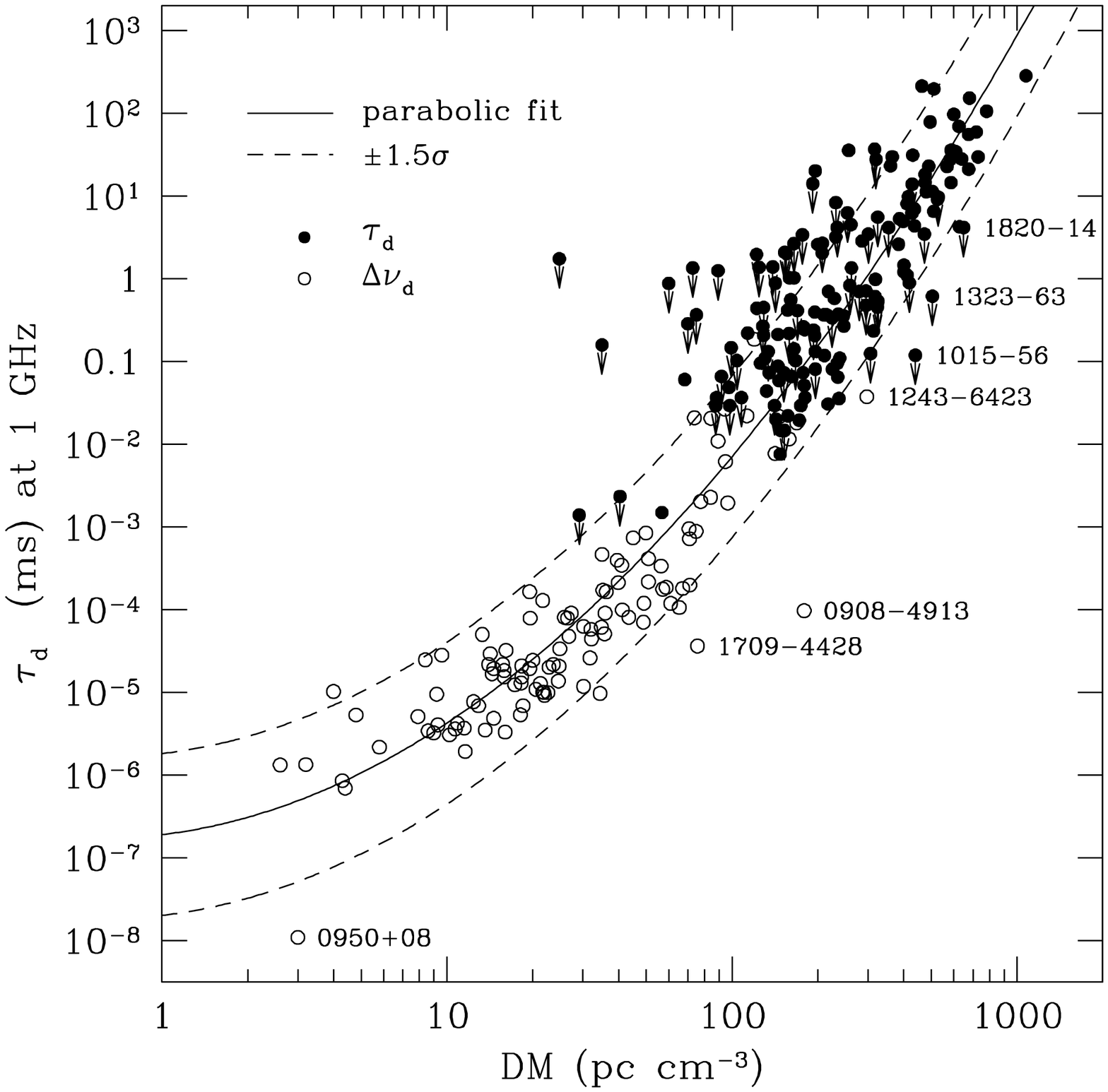}
\figcaption{
\label{fig:plottau}
Pulse broadening time at 1 GHz plotted against
dispersion measure.   Filled circles indicate direct measurement of
the scattering time; open circles represent scattering times estimated
from scintillation bandwidth measurements.  All measurements are
scaled to 1 GHz using $\tau_d\propto \nu^{-4.4}$.   The solid line
is the parabolic fit given in Eq.~\ref{eq:taudvsdm}; dashed lines are at
$\pm 1.5 \sigma$.
}
\medskip

Figure~\ref{fig:smvsdm} shows that the derived values of scattering
measure, when plotted against \DM, generally mimic the trend of pulse
broadening time against \DM\, though with significant differences
owing to the scatter of distance values for identical values of \DM.
The rough trend in Figure~\ref{fig:smvsdm} for $\DM \lesssim 20$ pc cm$^{-3}$ is
is in accord with what is expected for a homogeneous medium (e.g. Kuzmin 2001).
However, for $\DM\ \gtrsim 20$ pc cm$^{-3}$,
$\SM \propto \DM^3$ ,  a variation much stronger than the linear
trend expected for a (statistically) homogeneous medium in which 
\DM\ is strictly proportional to distance 
and $d\SM \propto d\DM$ (e.g. Shitov 1994).   
Though remarked upon by many as being anomalous and perhaps difficult to
explain  
the variation can be accounted for quite easily by relaxing
one or more assumptions that underly the calculation of \SM\ and its
interpretation.  In particular,  an inhomogeneous medium comprising
clumps of enhanced scattering with a power-law distribution of 
column densities, can account for the variations 
(e.g. Cordes, Weisberg \& Boriakoff 1985).

Following Cordes \etal\ (1991) and using Equations~11--12 of Paper~I,
we write
\be
\frac{d\SM}{d\DM} = 
	\csm F n_e,
\ee     
where $n_e$ is the local electron density.
The derivative is constant for a statistically homogeneous medium.
Recall that $\csm$ is a numerical constant that depends on the slope
of the wavenumber spectrum (Paper I) while the fluctuation parameter
$F$ depends on the outer scale, filling factor, and the local
ratio, $\dne/ n_e$.   
A non-Kolmogorov
wavenumber spectrum, combined with use of a Kolmogorov  spectrum
in calculating \SM, could cause seemingly anomalous values of
$d\SM/d\DM$ (Cordes, Weisberg \& Boriakoff 1985).   
Though evidence exists for non-Kolmogorov wavenumber
spectra along specific lines of sight, our assessment is that the
number of departures from such spectra is
not sufficient to produce the large variations and steep trend of
\SM\ with \DM.   

Systematic variations of the product $F n_e$ with location in the
Galaxy can easily account for the distribution of points in 
Figure~\ref{fig:smvsdm}.   However, large scale variations in the local mean
electron density alone are not sufficient.   In the large-scale
components of NE2001 (i.e., excluding clumps and voids), 
the local mean density changes by only a factor of 10.    Evidently,
systematic changes in $F$ are also needed.     
Evidence exists that the outer scale 
varies, being smaller in directions of intense scattering,
such as the lines of sight toward Sgr A* (Lazio \& Cordes 1998a, 1998b),
NGC6334B \cite{1990ApJ...348..147M}, and Cyg~X-3 (Wilkinson, Narayan,
\& Spencer~1994) and potentially larger in regions
of less intense scattering.  One also expects differences in 
filling factor and perhaps $\dne/n_e$ between the inner Galaxy
and the solar vicinity.   Therefore, the steep trend of \SM\
with \DM\ and the large fluctuations of \SM\ reflect a 
combination of smooth changes in electron density and fluctuation parameter,
and clumpiness of the ionized medium.
 
\medskip
\epsfxsize=8truecm
\epsfbox{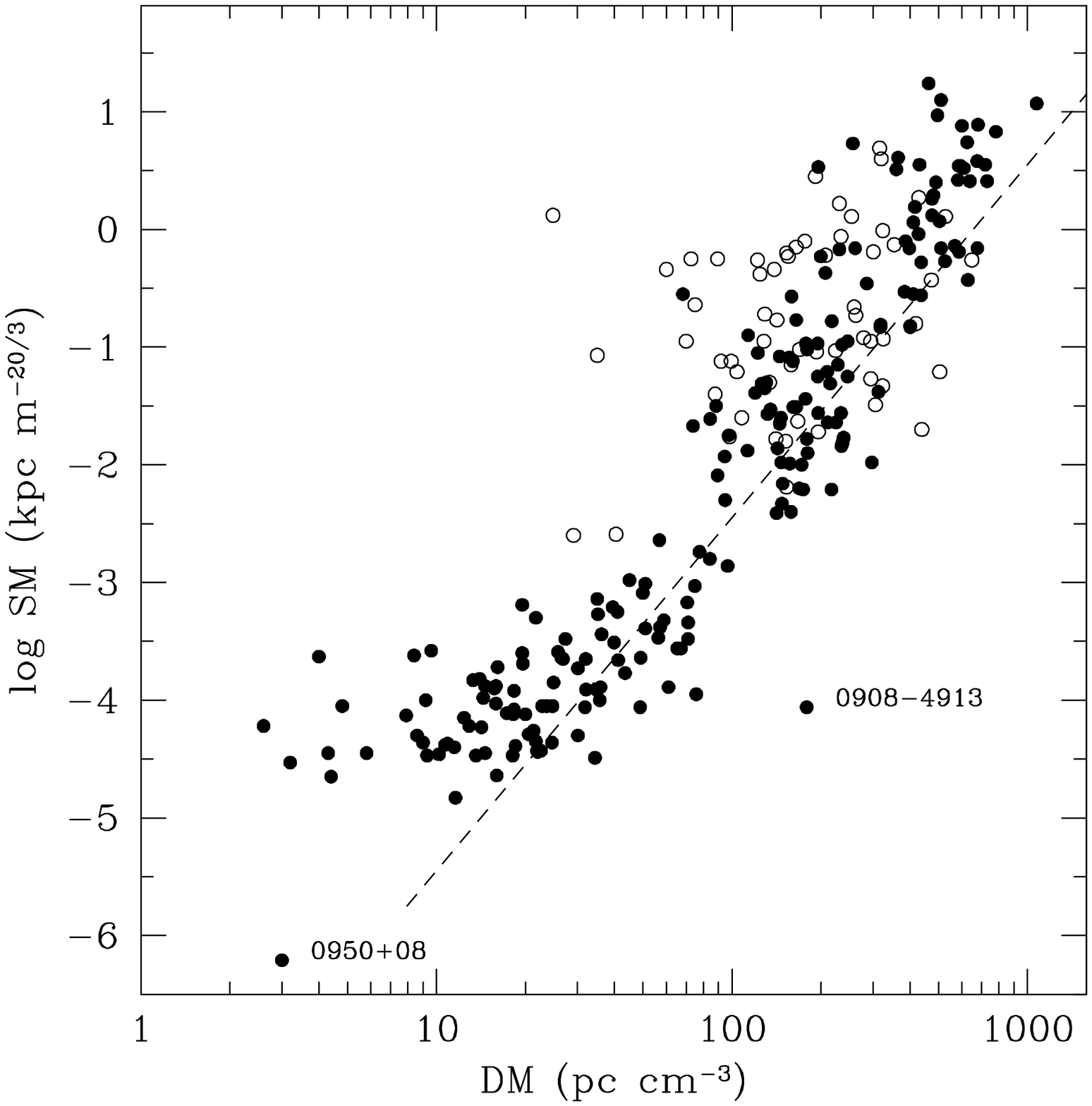}
\figcaption{
\label{fig:smvsdm}
\SM\ plotted against \DM\  for 269 pulsars.   Filled circles show measurements
while open circles are upper limits.  
The dashed line has slope 3 and is  placed by eye to indicate the
general trend of points for large values of DM.
The point near 
$\DM \approx 179$ pc cm$^{-3}$ and $\log \SM\ \approx -4$ 
is from a scintillation bandwidth measurement  
(Johnston, Nicastro \& Koribalski 1998)
of PSR J0908-4913,  whose line of sight passes  near or through
 the Gum Nebula  and Vela supernova 
remnant.   The low value of \SM\ is not caused by an overestimated distance
but, rather, by the low, effective value of the pulse broadening
time, as labelled in Figure~\ref{fig:plottau}. 
The same holds for pulsar B0950+08.
}

\section{Galactic Model for Electron Density}\label{sec:model}

As in TC93, we use a right-handed coordinate system 
$\xvec = (x,y,z)$
with origin
at the Galactic center, $x$ axis directed parallel to $l=90\arcdeg$,
and $y$ axis pointed toward $l=180\arcdeg$.  
The Galactocentric distance projected onto the plane  is
$r=(x^2+y^2)^{1/2}$. 

The structure of the model is described in Table 2 of Paper I and
associated text and includes thin-and-thick disk (axisymmetric)
components;  a local ISM component; a Galactic center component;
spiral arms; clumps of enhanced electron density or fluctuation
parameter; and voids with small electron density and fluctuation parameter.

\subsection{Modeling Issues}\label{sec:issues}

Our ultimate  goal is to estimate local values of mean electron density and
$\cnsq$ from line-of-sight measurements.   As with many 
inversion problems with insufficient sampling, significant
background information must be used to provide structure for
the model and, regardless, the result will not be unique.  This
is especially true for modeling the warm ionized medium (WIM)
in the Galaxy because
structure is known to exist on a wide range of scales 
that is  grossly undersampled by the available lines of sight. 
Our goal may be stated more realistically as aiming to model
the measurements (as integrated measures) for the available lines of sight 
and to extrapolate to other lines of sight for predictive purposes. 
As byproducts, we hope to identify and model Galactic structure in 
the WIM.

To illustrate issues of nonuniqueness, consider a single pulsar line of sight on which we have
measurements of \DM\ and pulse broadening, $\taud$, and perhaps constraints
on the range of possible distances, $[\dl, \du]$.   Suppose that
we model the line of sight with the simplest possible medium,
namely, a single \ion{H}{2} region or ``clump.''    The clump would
be characterized by $\DM_c$ and $\SM_c$ related to  the physical quantities 
$\nec, \Fc$, $\Delta s$, and $\dc$ (electron density inside the clump, 
fluctuation parameter,\footnote{%
The meaning of $\Fc$ is likely different for a single clump than for a
distributed medium as we originally define $F$ in Eq. 13 of Paper I,  
$F = \zeta\epsilon^2\eta^{-1} \ell_0^{-2/3}$. 
In the latter case,  $F$ includes factors involving the filling factor
($\eta$), the internal fractional variation of electron density,
$\epsilon = {\rm rms} n_e / n_e$, cloud to cloud variance ($\zeta$) of
clouds, and the outer scale, $\ell_0$.   For a single cloud, one may
consider $\eta \to 1$ and $\zeta \to 1$.
}
path length through clump, and distance from Earth),
\be
\DM_c &=& \nec\Delta s \\
\SM_c &=& \csm~\Fc~\nec^2 \Delta s. 
\label{eq:smclump1}
\ee
The pulse broadening yields a weighted \SM\ 
$\propto(\dc/D)(1-\dc/D)\SM_c$  
(cf.\ Eq. 7 of Paper I).
At best we have three measurements
(\DM, $\taud$ and $D\in[\dl,\du])$ and four unknowns,
$\DM_c, \SM_c, D$ and $\dc$.  Physically, $\DM_c$ and $\SM_c$
map into three unknowns, so there are really five total unknowns
($\nec, \Delta s,  \Fc, \dc$ and $D$) and at most three measurements.
If the same region is probed by multiple ($N_p$) pulsars and if the region
has simple, known geometry (so that $\Delta s$ can be calculated for
each LOS), then the number of unknowns is $N_p+4$ 
($N_p$  distances and four clump parameters) and there are $3N_p$ measurements  
if pulsar distances are known.   A unique solution exists for
$N_p\ge 2$ pulsars 
if observables are
known perfectly.   (With finite measurement errors,
the solution is unique in the mean.) However, if distances are not known, 
a unique solution requires $N_p \ge 4$ pulsars.  If 
$N_c$ clumps contribute to the observations of $N_p$ pulsars,   
the number of pulsars needed is $2N_c$ and $4N_c$, respectively,
with and without distance constraints on the pulsars.

In practice, sampling of the same structure by multiple pulsars
is relevant for large-scale features in the Galaxy (e.g., disk and
spiral-arm components) but not, generally,  for individual \ion{H}{2} regions.    
The typical spacing between lines of sight is simply too large with
the available sample to model small scale features such as \ion{H}{2}
regions uniquely.   Underdense regions such as supershells
 tend to be larger than standard
\ion{H}{2} regions, so there is hope to probe them.  To the extent that
large-scale structures dominate observables, it is plausible that
a model can be fitted  accurately.  However, any imperfections of
the adopted structure  will be manifest as systematic errors
and, of course,  small scale structures will be poorly modeled.   
An exception to this statement is the local ISM which, because
of its proximity, is sampled by many lines of sight.

\subsection{ Alternative Approaches}
Different approaches for modeling $n_e$ are
possible.  The approach of this paper is to adopt smoothly varying,
large-scale components that are perturbed by small-scale regions
that are under-or-over dense.
By contrast, one could adopt a model made up
solely of clumps that would correspond to structures containing
the WIM phase component.  One could thus
{\it synthesize} the large-scale structure out of those clumps
required for modeling \DM\ and \SM.   While this approach is
attractive in that it would rely only on \DM\ and \SM\ data to
define Galactic structure, there are insufficient lines of sight
available at present to allow discrimination
between large and small scale components.    Future pulsar surveys
that cover the Galaxy to the same or greater depth than the
Parkes Multibeam survey \cite{2001MNRAS.328...17M} may yield sufficient measurements, especially
when combined with scattering measurements of a large
sample of extragalactic objects.    A variation on this theme
is to recognize that the fundamental structures comprising the WIM
are filled \ion{H}{2} regions and shells, particularly supershells, having
underdense interiors and overdense walls.   With sufficient lines of
sight and using prior information regarding locations of known shells
and \ion{H}{2} regions, one could synthesize a large scale model   
out of such structures.

With sufficient lines of sight through the Galaxy, the 
best approach for fitting the Galactic density is a function
of the WIM itself,    in particular whether a smooth distribution
or a distribution based on discrete regions
is the best starting model. 
The statistics of observables will be determined by
the  filling factor of overdense regions and the mean free path
for intersecting them.  These factors, in turn, are related
to the overall porosity of the WIM which is affected by
percolation of regions of hot phase (shells and supershells).
If the WIM is largely smooth with occasional over and underdense
regions, a fitting approach that begins with smooth components
and then identifies outliers is warranted.   However, if the medium
is essentially a complex of over and underdense regions,  given
adequate sampling (numbers of lines of sight), explicit modeling
of individual regions would be best.   We take this first approach 
because small scale structures are undersampled
by the available  lines of sight;
also, a smooth model
appears to fit many objects quite well, though it does require
significant augmentation by clumps and voids.

\subsection{Smooth, Large-Scale Components}\label{sec:smooth}
\subsubsection{Outer, Thick Disk Component} \label{sec:thick} \label{sec:outer} 
The outer, thick disk component is responsible for the DMs of
globular cluster pulsars and the low-frequency diameters of
high-latitude extragalactic sources (e.g., as inferred from
interplanetary scintillation measurements and low-frequency
VLBI).  In TC93 this component
was determined to have a scale height of roughly 1~kpc with a
Galactocentric radial scale length of roughly 20~kpc.  However, the
data available for TC93 did not allow a firm
constraint on the Galactocentric scale length; scale lengths as large
as 50~kpc were also allowed by the data.

Lazio \& Cordes~(1998a) measured the angular diameters of eleven
extragalactic sources toward the anticenter, and Lazio \&
Cordes~(1998b) augmented these observations with those used by TC93
and other measurements in the literature to improve the constraints on
the Galactocentric scale length.  A likelihood method was used to
constrain not only the scale length, but also to search for any
signature of a warp.  Their analysis favors an unwarped, non-flaring
disk with a scale height of 1~kpc (though this may reflect the
non-uniform and coarse coverage of the anticenter provided
by the available data).

The likelihood is maximized for a radial scale length of~15--20~kpc,
but the data cannot distinguish between a gradual decrease in the
electron density and a truncated distribution.  Lazio \&
Cordes~(1998b) favored a truncated one, in which the scattering is
associated with massive star formation, which is also truncated near
20~kpc.  A radial extent of 20~kpc is also comparable to the radial
extent of H$\alpha$ emission observed for nearby spiral galaxies.

The functional expression for the electron density outside the solar circle is
${n_e}_1(r) = {n_e}_1 g_1(r)$, where
\begin{equation}
g_1(r) = \left\{\begin{array}{ll}

	\cos(\pi r/2A_1)\big/\cos(\pi R_{\sun}/2A_1), & \mbox{$r \le A_1$}; \\
	0, & \mbox{$r > A_1$}.
	
		      \end{array}
\right .
\label{eqn:radial2}
\end{equation}

In this functional form chosen by Lazio \& Cordes~(1998b) for the
electron density in the outer disk, there is no additional scattering
at $r > A_1$ because $n_{\mathrm{e}} =
0$~cm${}^{-3}$ for $r > A_1$.  An equivalent model is one in which
there is ionized gas beyond $A_1$  but fluctuations are turned off:
 $F_1 = 0$ for $r > A_1$.  Such  truncation
could occur if the distribution of scattering agents decreased more
rapidly with $r$ than does $n_{\mathrm{e}}$.  Given the current
paucity of both pulsars and scattering measurements toward the
anticenter, we do not regard this distinction as important.
Additional data, particularly if one could compare the DMs of distant,
anticenter pulsars with the scattering measures of extragalactic
sources, might allow this distinction to be resolved.

\subsubsection{Inner, Thin Disk Component} \label{sec:thin} \label{sec:inner}

An inner Galaxy component ($n_2$) consisted of a Gaussian annulus in
TC93.  Data available to TC93 could not distinguish a filled Gaussian
form in Galactocentric radius from an annular form, but the latter was
chosen for consistency with the molecular ring seen in CO (e.g., Dame
et al.~1987).  Here, we consider two alternatives.  First is the
annular form of TC93.  Second is a model based on a 99~GHz CS ($2 \to
1$) survey for ultra-compact \ion{H}{2} regions (which is assumed to
trace the OB star distribution) by Bronfman \etal\ (2000).

Bronfman et al.~(2000) report the surface density and scale height of
ultra-compact \ion{H}{2} (UCHII) regions, assuming an axisymmetric
distribution.  We have fit functional forms described below to these
quantities and determined the best-fitting parameter values using a
chi-square test.

The radial distribution of ultra-compact \ion{H}{2} regions peaks near 
0.7\rsun, with an approximately gaussian increase interior to
0.7\rsun\ and an approximately exponential decrease exterior to
0.7\rsun.  The scale length for the gaussian increase is 0.22~kpc
while the scale length for the exponential increase is 0.27~kpc.  The
$z$ distribution of ultra-compact \ion{H}{2} regions has a half-width
that is approximately constant at~30~pc interior to a Galactocentric
radius of 0.5\rsun, increasing through 100~pc by \rsun, and exceeding
200~pc by 1.5\rsun.  The $z$ distribution was modelled as a gaussian
with a scale height of~1~kpc.
In comparing the two models, we find that the second is not an
adequate replacement for an annular disk.  Our assessment is that
the UCHII population most likely resides in a spiral arm distribution
and may reflect a low-scale height component of the spiral arms 
(see below).   At the present time, we find that the annular component
combined with our spiral arm components provides the best fit.
In the future, the UCHII distribution  may be required when sufficient
lines of sight are probed that one or more is likely to pierce
an \hbox{UCHII}.


\subsubsection{Spiral Arms}

The spiral arm components in TC93 were modeled by specifying the locations
of the spiral arm centroids in the $(x,y)$ plane and then fitting for
the density, thickness and scale height of each arm.
Arm centroids were defined using
the locations of \ion{H}{2} regions augmented by use of directions where
radio continuum and neutral hydrogen (21~cm \ion{H}{1}) emission are enhanced
along spiral arm tangents.   We have reconsidered these definitions
because 
(a) recently discovered pulsars appear to require thicker arms 
because the TC93 model 
provides too few electrons in some directions to account for
measured DMs and it either over-or-under
predicts the scattering;\footnote{A thicker or more dense
spiral arm can either increase or decrease the scattering through
the competing effects of increasing the angular scattering while
bringing the pulsar nearer to the observer and also changing 
lever-arm effects that influence scattering observables.}
(b) there is evidence for the influence of a local (`Orion-Cygnus')
arm on the scattering of some pulsars (Gupta \etal\ 1994);
(c) the spiral arms in TC93 were truncated arbitrarily in their
arc lengths  and there is
evidence suggesting that they need to be extrapolated further; 
(d) on general principles, it is worthwhile to consider alternative,
mathematically defined forms for the spiral arms, such as the
logarithmic spiral forms considered in discussions of magnetic fields
in the Galaxy and in other galaxies (e.g., Vall\'ee~2002; Beck \etal~1996; Wainscoat
\etal\, 1992).

The functional form of the spiral arm density is
\be
\nearms &=& n_a G_a(\xvec), \\
G_a(x,y,z) &=& \sum_{j=1}^{\narms} f_j\, {g_a}_j(r,s_j/w_j)\, h(z/h_j h_a), 
\ee
where the summation ranges over $\narms = 5$ spiral arms.  
Factors of order unity, $f_j, w_j$ and $h_j$, allow control of the
electron density, width and scale height of each spiral  arm, respectively.
Each arm also has a separate $F$ parameter, ${F_a}_j, j=1,5$.

Figure~\ref{fig:armdefs} shows the locations of spiral arms
as defined in TC93 and as modified by us.
For the most part, the spiral arm definitions in TC93 are
satisfactory to account for the overall shape of the
distribution of \DM\, vs. $\ell$.   The fast rolloff of \DM\, 
for negative longitudes at $\ell\sim -75\arcdeg\pm 15\arcdeg$ 
is associated with the ``Carina'' side of the Carina-Sagittarius arm
while the corresponding ``Sagittarius'' rolloff is at positive longitudes
at $\ell\sim 45\arcdeg\pm 5\arcdeg$.   Analogous features from
arms interior to the Carina-Sagittarius arm are not obvious 
in the figure owing to crowding of objects.  However, the local
``Cygnus-Orion'' arm and possibly the exterior
``Perseus'' arm are more evident at positive than negative longitudes
because they are nearer the Galactic center at positive longitudes
and appear to have greater densities as a consequence.  This asymmetry
is manifested in Figure~\ref{fig:lbplot8} by the slower rolloff
of \DM\, for $60\arcdeg \lesssim \ell \lesssim 180\arcdeg$ than in the
corresponding range of negative longitudes.   

The spiral arm centroids are defined as perturbed logarithmic spirals.
To begin, we define the centroid of the $j^{\mathrm{th}}$ arm according to
\be
\theta_j(r) = a_j \ln(r/r_{{\rm min},j}) + \theta_{{\rm min},j},
\ee
where $\theta_j$ is measured anticlockwise
from the $y$ axis (with the Galactic center
at the origin increasing positively toward the Sun) and $r$ is
Galactocentric radius.    Values for $a_j$, $\theta_{{\rm min},j}$
and $r_{{\rm min},j}$ are taken from Wainscoat \etal\ (1992).  
The spiral structure is similar to that proposed by Ghosh \& Rao (1992)
and also described by Vall\'ee (1992).   We use a Sun-Galactic Center
distance of 8.5 kpc, the IAU recommended value (but see \S\ref{sec:model.discussion}).
We perturb the shapes of the arms  in the vicinity of the Sun
to match the shapes defined in TC93.  These perturbations are guided
by the detailed distributions of \DM\ in Galactic longitude and
by the locations of spiral arm tangents in a number of tracers.
In practice, we tested three spiral arm models against the data:
(1) the TC93 spiral arms;
(2) pure logarithmic spirals as described in Wainscoat \etal\ (1992);
and
(3) logarithmic spiral arms perturbed to match arm shapes near the
Sun as described in TC93.     The likelihood analysis disfavors
the first case and slightly prefers model (3) over model (2).

{\it Caveat:} The spiral arm components in our model, like
those in TC93, are modeled as overdense regions.  Astrophysically, 
however, the enhanced star formation in spiral arms will produce
underdensities  as well as overdensites, as has been demonstrated
by the identification of supershells.  Though we have
adopted spiral arms as overdense regions, to account for \DM\
and \SM\ we have had to introduce  ellipsoidal underdense perturbations
(``voids'') in particular directions, as discussed below.     

\subsection {Local ISM (LISM)}\label{sec:lism}

The local ISM displays density enhancements and deficits
associated with the particular star-formation history,
including supernovae, in the region  and with existing stars
that ionize local gas.  As such, the LISM is not at all unique
in the Galaxy.  However, we must model it carefully because we
view the rest of the Galaxy --- and the entire Universe ---
through it.

We model the local ISM in accord with \DM\ and \SM\ measurements of
nearby pulsars combined with parallax measurements  and guided by 
\halpha\ observations that provide estimates of \EM.  
Observations and analysis by Heiles (1998)
and Toscano \etal\ (1999 and references therein)
suggest the presence of four regions of low density near the Sun:
(1) a local hot bubble (LHB) centered on the Sun's location that is
	long known because of its prominence in \ion{H}{1} and X-ray observations;
(2) the Loop I component (North Polar Spur) that is long known because
	of its prominence in nonthermal continuum maps;
(3) a local superbubble (LSB) in the third quadrant; and
(4) a low density region (LDR) in the first quadrant.
Additional features have been identified by Heiles (1998) but the available
lines of sight to pulsars appear to not require their inclusion in our
model. 
Bhat \etal\ (1999) explicitly fitted for parameters of the LHB
using pulsar measurements and a model having a low-density structure 
surrounded by a  shell of material that produces excess scattering.
Some of the parallax distances used by Toscano \etal~(1999)
have been revised (Brisken \etal\ 2000; Brisken 2001; Brisken \etal\ 2002),
in some cases substantially, implying lower densities
in the third quadrant than they inferred. 

We have used the work of Heiles~(1998),  Bhat \etal~(1999), and Toscano \etal~(1999), along with comprehensive work by Frisch (1996, 1998)  as  guides
for defining structures in the local ISM but we have made
new fits to the locations, sizes, shapes and densities in order
to estimate both the distances and the scattering of nearby pulsars
to within the measurement errors. 
We include the four structures as follows.  For the LSB and LDR, we use a
gaussian ellipsoid with uniform electron density and $F$ parameter 
inside the $1/e$ contour.     
For the Loop I structure, we use a spherical volume of low density surrounded
by a shell of higher density and fluctuation parameter.
The LHB has a more complex structure
as delineated by X-ray observations (Snowden \etal\ 1998)
and \ion{Na}{1} absorption (Ma{\'\i}z-Apell\'aniz~2001 and
references therein).   Based on contours shown in 
Sfeir \etal\ (1999), we model the LHB as a cylinder with
ellipsoidal cross section having constant area for $z\ge0$
and decreasing area for $z<0$.   The cylinder is vertical in the
$x$-$z$ plane and slanted in the $y$-$z$ plane such that the cylinder
axis has $dy/dz >0$.    As suggested by Sfeir \etal~(1999) and by
the results of our fitting, the LHB has a ``blow out'' structure 
for positive $z$, but is pinched off at negative $z$.
Appendix A describes the mathematical models used for the four regions.  
Table 4 
of Paper I lists the parameters of the local ISM model
and gives values based on the fitting we describe below.

Because $n_e$ is smaller inside each
of these regions than outside, we define a weight factor~$\wlism$ that is unity for a location inside any of the four regions
and is zero outside.   As seen in Table 1 of Paper I, 
the electron density is set to either the LISM value or the ambient value
by using $\wlism$ as a switch.   Similarly, because the LHB
has lower density and $F$ parameter than the LDR and LSB, we assign a weight to
the LHB component, $\wlhb$, that is unity inside its $1/e$ contour
and which allows the LHB component to determine $n_e$ and turns off
any contribution from the LSB and LDR components.  We also use the
weight for the LSB component to override the LDR component.   The
LISM model therefore is of the form
\be
\nelism(\xvec) &=& (1-\wlhb)  \{ (1-\wloopI)  \times \nonumber \\
	       && 
	[ (1-\wlsb)\neldr(\xvec) + \wlsb \nelsb(\xvec) ] + \nonumber\\
               &+& \wloopI \neloopI(\xvec) \} + \wlhb \nelhb(\xvec) 
\label{eq:lism}
\\
\Flism(\xvec) &=& (1-\wlhb) \{ (1-\wloopI) \times \nonumber \\
	&& 
	[ (1-\wlsb)\Fldr(\xvec) + \wlsb \Flsb(\xvec) ] \nonumber \\ 
               &+& \wloopI\FloopI(\xvec) \} + \wlhb \Flhb(\xvec) \\
\nonumber \\ 
\wlism(\xvec) &=& {\rm max}(\wldr, \wlhb, \wlsb, \wloopI).
\ee

\subsection{Gum Nebula and Vela Supernova Remnant} \label{sec:gum}

Features in or near 
the nearby Gum Nebula have discernible influence on the dispersion measures
and scattering measures of several pulsars.  In TC93 a large region
was included which perturbed the dispersion measures of pulsars viewed 
through the Gum Nebula but did not influence the scattering. 
This choice was made because, at the time, there were insufficient lines
of sight with pulse broadening measurements
 to model scattering of the Gum Nebula.  
Since the writing of TC93,  investigations of the Vela pulsar and other
pulsars indicate that scattering is large within a region
of at least 16 degrees diameter centered roughly on the direction
of the Vela pulsar  (Mitra \& Ramachandran 2001).   We have modeled
the Gum/Vela region based on the scattering measurements and also
on the fact that an enhanced, local mean electron density is required
to account for the dispersion measures of five objects,

We model the region with an overlapping pair of spherical regions,
one describing the Gum Nebula, the other the immediate vicinity of
the Vela pulsar. 

The pulsars with high DMs and overestimated distances are
     B0736$-$40, J0831$-$4406, B0808$-$47, B0833$-$45 (Vela), 
     B0743$-$53, J0855$-$4658, and B0950$-$38. 
For all objects except Vela,  the model DM without the Gum/Vela components
 is smaller than the measured DM even when integrated to 30 kpc.  
For Vela the estimated distance is 2.8 kpc without these components,
compared to recent distance estimates, $0.25 \pm 0.03$ kpc
(Cha \etal~1999) and $0.29^{+0.076}_{-0.05}$ kpc (Caraveo \etal\ 2001).
 


\subsection{Galactic Center (GC) Component}\label{sec:gc}

The scattering diameters of \sgra\ and several nearby OH masers
indicate that a region of enhanced
scattering is along the line of sight to the Galactic center
(van Langevelde \etal\ 1992; Rogers \etal\ 1994; Frail \etal\ 1994). 
Typical diameters, scaled to~1~GHz, are approximately 1\arcsec, roughly 10 times
greater than that predicted by TC93, even with the general enhancement
of scattering toward the inner Galaxy in that model.

Scattering measurements of GC sources alone do
not constrain the \emph{radial} location of the
scattering region for the following reason: 
for such sources, a region of
moderate scattering located far from the Galactic center can produce
angular broadening equivalent to that from a region of intense
scattering located close to the Galactic center.  Indeed, previous estimates
for the location of the scattering region ranged from~10~pc to~3~kpc
from the GC.

The degeneracy between the scattering strength and radial location can
be broken by observing distant sources through the scattering region.
Lazio \& Cordes~(1998c) conducted a survey for extragalactic sources
observed through the Galactic center.  They found a paucity of
sources, suggestive of a hyperstrong scattering region located in or
near the Galactic center.  Lazio \& Cordes~(1998d) combined the
results of this survey with the extant radio-wave scattering data and
free-free emission and absorption measurements in a likelihood
analysis that constrained the GC-scattering region separation,
$\delgc$, and the angular extent of the region, $\psi_\ell$.  The
maximum likelihood estimates of these parameters were $\delgc =
133_{-80}^{+200}$~pc, $0.5\arcdeg \le \psi_\ell \lesssim 1\arcdeg$.
Lazio \& Cordes~(1998d) argued that the close correspondence between
$\delgc$ and $\psi_\ell\dgc$ indicates that the scattering region
encloses the \hbox{GC}.

Lazio \& Cordes~(1998d) suggested an axisymmetric Gaussian
form for the GC region with radial scale $\RGC=0.15$ kpc 
and scale height $\HGC=0.075$ kpc with a 
nominal central density $n_{\mathrm{GC}} = 10$~cm${}^{-3}$, which they 
estimated using free-free absorption measurements incorporated
into their likelihood analysis.  
We have altered this model in order to match the scattering diameters
of nearby sources.   In particular,  the data require a
smaller scale height $\HGC\approx 0.026$ kpc and slightly smaller
radial scale, $\RGC\approx 0.145$ kpc.  In addition we have offset
the center of the distribution by 
$(x_{\rm GC}, z_{\rm GC}, z_{\rm GC}) = (-0.01, 0, -0.02)$ kpc.
The model for the electron density in the GC is
\be
n_e(\xvec)  
 &=& \exp\left[
	 -\left( \frac{\delta\rperp^2}{\RGC^2} 
         + \frac{(z-z_{\rm GC})^2}{\HGC^2} 
	  \right)
       \right] \\ 
\delta\rperp^2 &=& (x-x_{\rm GC})^2 + (y - y_{\rm GC})^2,\nonumber
\ee
which is truncated to zero for arguments of the exponential 
smaller than $-1$.    The truncation allows the GC component to
leave unaffected lines of sight to nearby OH masers that show much
less scattering than those seen through the GC component. 
A value 
$F_{\rm GC} \sim 6\times 10^4$ produces SM values needed to account for the 
scattering diameters of \sgra\ and the OH masers.

\subsection{Regions of Intense Scattering (``Clumps'')}

In addition to regions that perturb the scattering of
the Vela pulsar and other pulsars near it and the Galactic center, other
regions of intense scattering must exist to account for the large
angular diameters and/or pulse broadening seen toward a number of
Galactic and extragalactic sources.

We define ``clumps'' as regions of enhanced $n_e$ or $F$, or both,
and identify them by iterating with preliminary fits to the smooth
components of the electron-density model (e.g., the thin and thick disk
components and the spiral arms).  

We model clumps with
thickness $\Delta s \ll D$ and we use parameters $\nec, \Fc$ and $\dc$
(electron density inside the clump, 
fluctuation parameter, and distance from
Earth).  The implied increments in \DM\ and \SM\ are
\be
\DM_c &=& \nec\Delta s \nonumber \\
\SM_c &=& \csm~\Fc~\nec^2 \Delta s. \nonumber \\
      &=& 10^{-5.55} \DM_c^2 F_c / \Delta s_{kpc},
\label{eq:smclump}
\ee
where the last equality holds for \DM\ and \SM\ in standard units.
Eq.~\ref{eq:smclump} implies that relatively modest contributions
to \DM\ can produce large changes in \SM\ if the clump is small and 
fluctuation parameter large.  For example, 
$\DM_c = 10$ pc cm$^{-3}$, $F_c = 1$ and $\Delta s = 0.02$ kpc
yield $\SM_c = 0.014$ kpc m$^{-20/3}$.   For a pulsar 1 kpc away,
the clump would perturb \DM\ and  the resultant
distance estimate by perhaps 30\% while \SM\ would increase by about 
a factor of 100.   Thus pulsars which have anomalous scattering
relative to the smooth model can, in many cases,
 be well modeled with small perturbations to the \DM\ predictions
of the model.  This also means that the model is expected to
be much better for distance and \DM\ estimation than for
scattering predictions.     

In some cases the pulsar's distance is much smaller than implied
by DM and the smooth parts of the model.  If the pulsar's distance
is well constrained, then we require $d_c < \dl$ and $\DM_c$ is
bounded by the requirement that the new distance estimate
(including the clump) must match the measured distance to within
the errors.

\subsection{Regions of Low Density (``Voids'')}\label{sec:voids}

We also found that some pulsar distance constraints could not be satisfied
using the previously defined structures without recourse to placement
of a low-density region along the line of sight.  We call these
regions ``voids'' although they simply represent, typically,
lower-than-ambient density regions.  By necessity, they take precedence
over all other components (except clumps), which we effect by usage of a void
weight parameter, $\wvoids = 0,1$, that operates similarly to
$\wlism$.   The mathematical form is given in Table~2 of \hbox{Paper~I}.
We use elliptical gaussian functions with semi-major and semi-minor
axes $a, b, c$ and rotation angles $\theta_y, \theta_z$ about the
$x$ and $z$ axes.

\subsection{Components Not Included}

We tested inclusion of a Galactic bar component using 
a shape guided by the work of Blitz \& Spergel (1991)
and others, but varied the size, orientation
and density.   We found that the likelihood function was not 
increased sufficiently by inclusion of the bar to warrent inclusion
in the final model.  This insensitivity to a bar component may arise
from an insufficient number of pulsars that probe it and also
because of degeneracy with the inner ring and inner spiral-arm 
components.

\section{Methodology of Parameter Estimation} \label{sec:methodology}

The electron density model is a combination of large and small
scale structures.  Some components are sufficiently disjoint
from others that their parameters can
be estimated without a global analysis.   
The Galactic center component is the best example of such independence,
and we have used the model described in Lazio \& Cordes (1998d), with
the modifications described in \S\ref{sec:gc}.
The local ISM structures are independent of many, but not all,
of the large-scale components. 
However,
other components are highly coupled. Given that the nonlinear model
contains too many parameters to allow a straightforward
optimization,  we have taken an iterative, ad hoc approach
that yields an acceptable solution.  However, we emphasize that
the model is not unique, and it is probable that we have not found
the best, global  solution for the assumed structure of the model.   
Additionally, we have included only those structures required by
the dispersion and scattering data.  Alternatively, we could have also 
included known \ion{H}{2} regions whose sizes and densities are well estimated.
At present, most known pulsars are unaffected by
 known \ion{H}{2} regions.   In the future, when more pulsars are known, inclusion
of \ion{H}{2} regions will become warranted.

\subsection{The Role of the Galactic Pulsar Distribution}
\label{sec:role}

The optimal procedure would consist of simultaneous  modeling of the Galactic 
distributions of pulsars and the electron density  (for the case where
the distances of most pulsars are not known independently).   
Such a procedure requires accurate knowledge of the pulsar luminosity
function which, owing to beaming of the pulsar radiation, is not
yet known (but see Arzoumanian, Chernoff, \& Cordes~2002).   
For our present effort, we assume that there is considerable spatial overlap of
Galactic pulsars and electrons.   In fact, we assume that the overlap
is such that the vast majority of pulsar distances are {\it determinable}
from their dispersion measures.  Counterexamples exist, of course,
in the pulsars within globular clusters that are well outside the Galactic 
plane  whose DM-determined distances have large errors.     Objects 
in the LMC and SMC also have distances indeterminable from the
measured DM. For most other objects in the known pulsar sample,
however we assume that they are near enough so that 
$\DM < \DMinfty$.  Thus, we impose on our model the requirement that
$\nover = 0$ for objects not in globular clusters or the Magellanic
clouds (see below).  This requirement may bias the model in some directions because
high-velocity objects are expected that, over a typical pulsar lifetime
$\sim 10$ Myr, can move well outside the regions of significant free electron
density. 

\subsection{Component Inclusion and Figures of Merit}
\label{sec:components}

Our overall fitting philosophy follows Occam's Razor in that we attempt
to account for the dispersion and scattering of as many lines of sight
as possible using large scale structures.  Only when required do we
include, ad hoc,  an additional small-scale perturbing cloud or clump
along a particular line of sight.  As diagnostics and  
figures of merit for the model,
we used:
\begin{enumerate}
\item The likelihood function for pulsar distances and DMs,
$\lpsrd$, is maximized as one measure of best fit to the available
measurements.  
\item $\nhits$, the number of pulsars having independent distance 
measurements, expressed as a range $[\dl, \du]$, and for which
the model yields $\dhat \in [\dl, \du]$.  We maximize $\nhits$.
Maximizing $\nhits$ is identical to maximizing the likelihood function
for distance and DM data, $\lpsrd$, defined below.    However,
in practice, we tended to use $\nhits$ for constraining the local
ISM and $\lpsrd$ for the larger scale medium.
\item $\nover$, the number of pulsars with  
$\DM > \DM_{\infty}(\ell, b)$, the maximum \DM\ allowed by the model
when integrating to infinite distance. We minimize $\nover$.
When fitting large-scale components to account for pulsar DMs,
we calculated a DM-residual plot for those pulsars in a given
iteration that had DMs larger than the model could account for.
By plotting $\Delta\DM = \DM - \DM_{\infty}$ against Galactic
coordinates $\ell, b$, we identified which model components needed
adjustment.   For the TC93 model,  a plot of such residuals is shown 
in Figure 4 of \hbox{Paper~I}.  
\item $\nlum$, the number of pulsars with discrepant luminosities,
where ``luminosity'' is defined as the pseudo-luminosity,
$\Lp = S \dhat^2$, where $S$ is the period-averaged flux density
(typically at either 0.4 or 1.4 GHz).  The goal is to minimize $\nlum$.   
In practice, utilizing $\nlum$ is subjective for individual objects
because pulsar surveys are subject to Malmquist-type bias that is
complicated by the fact that both the 
detection limits and the pseudo-luminosities are period and \DM\ dependent.
Nonetheless, one expects that the upper envelope on 
$\Lp$ should be constant as a function of $\DM$ if there is a cutoff
in the luminosity function and once sufficient volume
has been searched to find the most luminous objects.
For the present work, we have only made cursory use of
this figure of merit because use of $\nover$ accomplished essentially
the same thing.
\item Likelihood functions for scattering  observables as defined 
below: $\lpsrs$, $\lgals$ and $\lxgals$ for the separate
subsamples of pulsars, other Galactic sources, and extragalactic sources.
As usual, we maximize these likelhoods.
\end{enumerate}



\subsection{Definitions of Likelihood Functions}\label{sec:like2}

We calculate the likelihood function 
using model predictions and 
PDFs for the errors in the observables.
The likelihood function is factorable
according to the statistical independence of four kinds of
measurements: 
(1) pulsars with independent distance constraints;
(2) scattering measurements of pulsars (pulse broadening or
    scintillation bandwidth);
(3) scattering measurements of other Galactic sources;
and
(4) scattering measurements of extragalactic sources.  
We factor the total likelihood, $\like$, into a factor for
pulsar distances and a factor for all scattering measurements:
\be
\like &=& \lpsrd \ls \\
\ls   &=& \lpsrs  \lgals  \lxgals. 
\ee
For the distance factor, we use the fact that some pulsars are
attributed distances in an interval $[\dl, \du]$
with essentially no preference for any distance within the interval.
We therefore adopt a flat distribution inside the interval
and exponential decays for model distances outside the interval.
The e-folding scales are proportional to $\dl$ and $\du$. Thus
the distance PDF is
\be
f_{\rm D}(D; \dl, \du) = \norm \times
\left \{ 
\begin{array}{ll}
  1	 			& \dl \le D \le \du \\
\\
e^{-(\dl - D)/\epsilon\dl} 	& D<\dl \\
\\
e^{-(D - \du)/\epsilon\du} 	& D>\du,
\end{array}
\right.
\ee 
where 
\be
\norm = \left [
	\du - \dl + \epsilon(\dl + \du - \dl e^{-1/\epsilon})
	\right ]^{-1}.
\ee
We choose $\epsilon = 0.05$ because distance constraints are
typically hard bounds for \ion{H}{1} absorption, as opposed to describing
Gaussian error bounds.  The likelihood factor for distances
becomes
\be
\lpsrd = \displaystyle\prod_{j=1}^{\npsrd} f_{\rm D}(\dhat_j; \dl, \du),
\ee
where $\dhat_j$ is the model prediction for a given set of
model parameters.\footnote{%
Alternatively, we could use \DM\, as the observable and then use
the empirical distance constraint ($D\in[\dl, \du]$) to calculate
a set of predicted values of \DM.     The two approaches are
equivalent and we choose to use distance as a quasi-observable
quantity.}

For scattering measurements we assume that logarithms of
observables $X = (\dnud, \taud, \theta_d)$ are normally
distributed with standard deviation $\sigma \approx \sigma_X/X \ll 1$.
There are also upper limits, particularly on $\taud$,
for which we use the cumulative distribution function of $\log X$ to
evaluate likelhood factors.  
The likelihood 
factors involving scattering observables are
\be
\like_{y,s} = {\prod_{j=1}^{N_y}} \like_j,
\ee
where y=(psr, gal, xgal) and $N_y$ is the number of measurements of
each kind and $j$ labels the relevant line of sight.
Model predictions are compared with the scattering observables, 
$\theta_d$, $\taud$ and $\dnud$, through the ratio 
$\rscatt = \widehat{\theta_d}/\theta_d$,
$\rscatt = \widehat{\taud}/\taud$,
or
$\rscatt = (\widehat{\dnud}/\dnud)^{-1}$, 
such that $r<1$ signifies that the model {\it underestimates} the scattering
strength along the particular line of sight.   
The likelihood factor for the j$^{th}$ line of sight is 
\be
\like_j = \left\{
\begin{array}{ll}
(2\pi\sigma^2)^{-1/2} e^{-(\log \rscatt)^2/2\sigma^2} & 
	\mbox{\rm measurements} \\
\\ 
\frac{1}{2}\left\{1 + {\rm erf}(\log \rscatt/\sqrt{2}\sigma) \right\} &
	\mbox{\rm upper  limits}.
\end{array}
\right.
\ee

\subsection{Procedure}

We took the following, general approach to fitting for model
parameters:
\begin{enumerate}
\item Adopt preliminary values of parameters based on
TC93 and on empirical distributions such as those shown in
Figures~\ref{fig:dmz}--\ref{fig:smvsb}.
\item 
\label{item:large-scale1}
Fit for parameters of the large-scale components
of $\negal$ both before and after identifying and excluding outlier points
that are associated with pecularities of specific lines of sight.
This fit is made by maximizing the likelihood functions
defined below.
\item 
\label{item:local1}
Fit for local ISM parameters by freezing the parameters 
of large scale structure   and using only pulsars 
with $\DM \le 40$ pc cm$^{-3}$ and for which 
parallax distance constraints exist.  Initially, only
distance and \DM\ data are used; then scattering measurements
are used as well.
\item 
\label{item:large-scale2}
Re-fit parameters of the large scale distribution by 
identifying the best, compromise model that nearly minimizes 
$\nover$ and  nearly maximizes $\nhits$
while achieving near-maximum likelihoods.

\item 
\label{item:local2}
Re-fit the local ISM parameters again while holding
fixed other parameters and maximizing $\nhits$.
\item 
\label{item:large-scale3}
Re-fit all large-scale parameters again while holding
fixed the local ISM parameters.
\end{enumerate}
Steps \ref{item:large-scale1}--\ref{item:large-scale3} were
taken first by fitting for those parameters that determine
\DM\ while ignoring scattering.  Thus, in a first pass,
we optimized those parameters that determine the local mean $n_e$.
The same steps were repeated in a second pass to optimize
the parameters that determine the local mean
$\cnsq$.  For this pass, we maximized the likelihood function
for scattering observables (defined below) and we maximized
the number of nearby objects with parallax distances 
for which the scattering observable was within an acceptable
range about the measured values.   Once the best, smooth
model was obtained in this way,  we identified ``outlier''
lines of sight which required ``clumps'' of excess
$n_e$ and/or $\cnsq$.     We modeled these clumps to
bring the outliers into conformance with measured quantities.
Finally, we repeated the whole procedure once the individual
clumps were included to determine the best parameters of
the smooth components of the model. 

\subsection{Identification of Regions of Enhanced Electron Density}

Once solutions for the large-scale components are chosen, clumps are included in the model
and a new fit is done for parameters of smooth components in the model.
The clump identification process is also iterated.  Our final solution
results after two such iterations.  Lines of sight  with required clumps
are given in Tables~5--7 of \hbox{Paper~I}.

We identify clumps empirically
using the following method.  Starting with an initial smooth
model for the Galactic $n_e$, which consists of $\negal$, $\nelism$,
and the Galactic center component, $\negc$, we investigate the line of
sight for each object for which we have a scattering measurement.
First consider a pulsar.   If the model fits the distance constraints
(if any) and the measured scattering observable to within a factor 
of two, no
additional electrons are needed along that line of sight.  If the
scattering is outside a factor of two and/or $\dhat> \dl$ then  a trial
clump is placed along the line of sight to bring relevant
estimates into agreement with the observables.     
For specificity, we adopt a clump with a
Gaussian density profile with radius $\rc$ whose centroid sits on
the line of sight. We choose $\rc = 10$ pc in most cases  as an arbitrary but
not unreasonable clump radius given that some \ion{H}{2} regions are of
order this size.  The clump distance (from the Sun) is chosen
as being either half the lower pulsar distance limit if meaningful distance
constraints exist, $\dc = \dl/2$, for nearby pulsars
or it is set to the midpoint of the spiral arm which contributes
the longest path length to the line of sight.   This latter choice
is clearly not unique for cases where the line of sight traverses
multiple spiral arms, but without further information, this seems the
most probable course to take.  Once $\dc$ is chosen,   we step
through a grid of values of electron density, $\nec$, 
and fluctuation parameter, $\Fc$, and find pairs of values that 
bring the overall model values into conformance with the data.

As discussed below, this procedure generally finds multiple solutions. 
We select the one solution that has the minimum values for
$\nec$ and $\Fc$ while allowing a good fit. 

Using a preliminary model, we calculate the ratio $\rscatt$ (defined above)
and identify lines of sight with significant departures from unity. 
Next we identify, where possible, clump parameters that bring prediction
and reality into agreement, i.e., $0.5 \le \rscatt \le 2.$ 

In assigning clumps, we  analyzed each line of sight separately.
We attempted to find multiple LOS affected by larger-scale clumps.
Though individual OH masers (e.g., those in W49) can be treated in this way,
we find that the density of known, scattered sources in the sky is too low
to generally find objects that are affected by the same 
clump or region of enhanced scattering.
We find
$\Fc = 0$ in some cases which causes the distance to lessen and thus
the scattering to change.  


\medskip
\epsfxsize=8truecm
\epsfbox{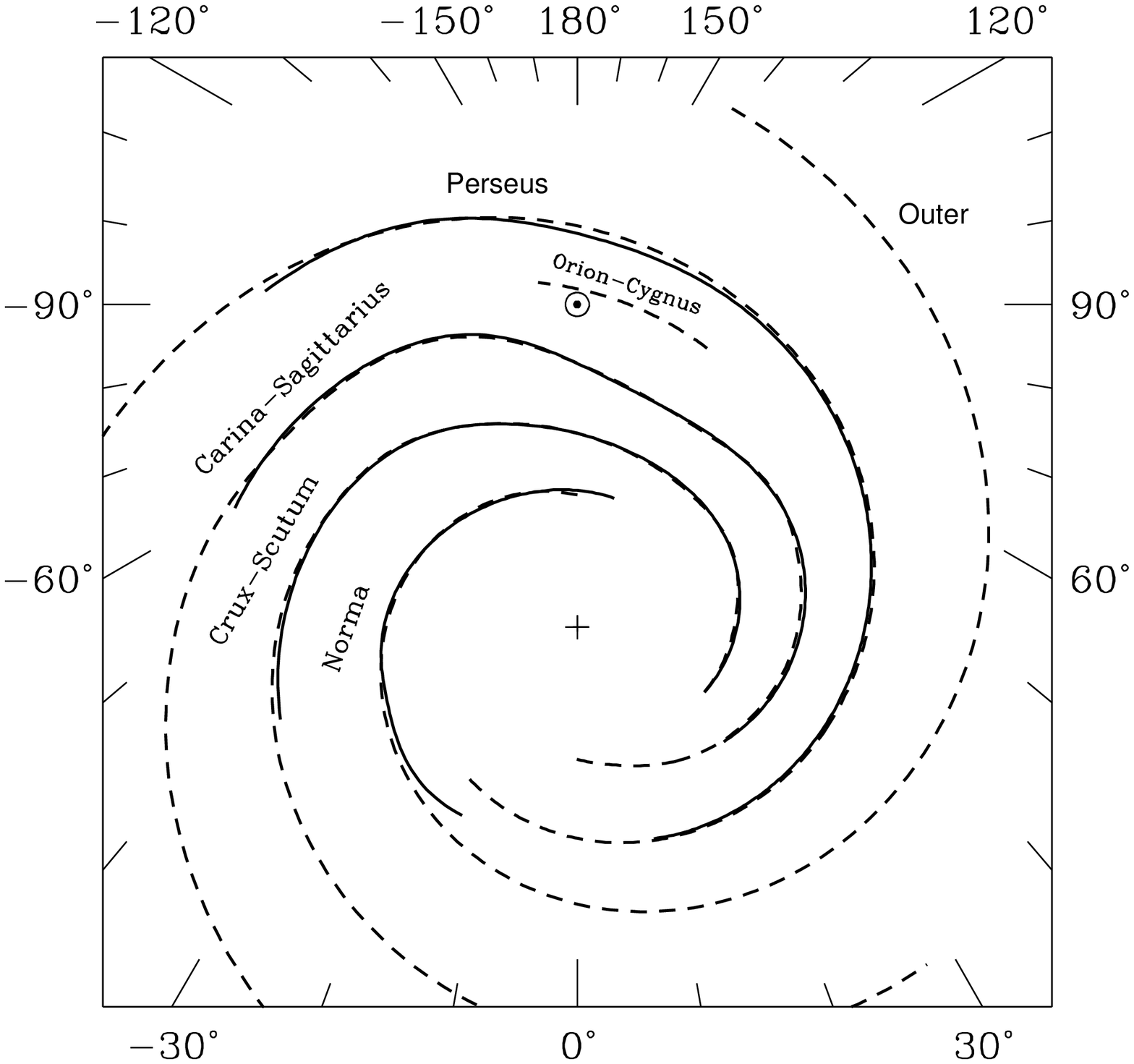}                
\figcaption{
\label{fig:armdefs}          
Solid lines: spiral model of the Galaxy used in TC93,
defined according to work by  Georgelin and
Georgelin (1976), modified as in TC93.
Dashed lines:  a four-arm logarithmic spiral model combined
with a local (to the Sun) arm using
parameters from Table 1 of Wainscoat \etal\ (1992),
but modified so that the arms match some of the features of
the arms defined in TC93.
The names of the spiral arms, as in the astronomical literature,
are given.
A $+$ sign marks the Galactic center and the Sun is
denoted by $\odot$. 
}

\medskip
\epsfxsize=8truecm
\epsfbox{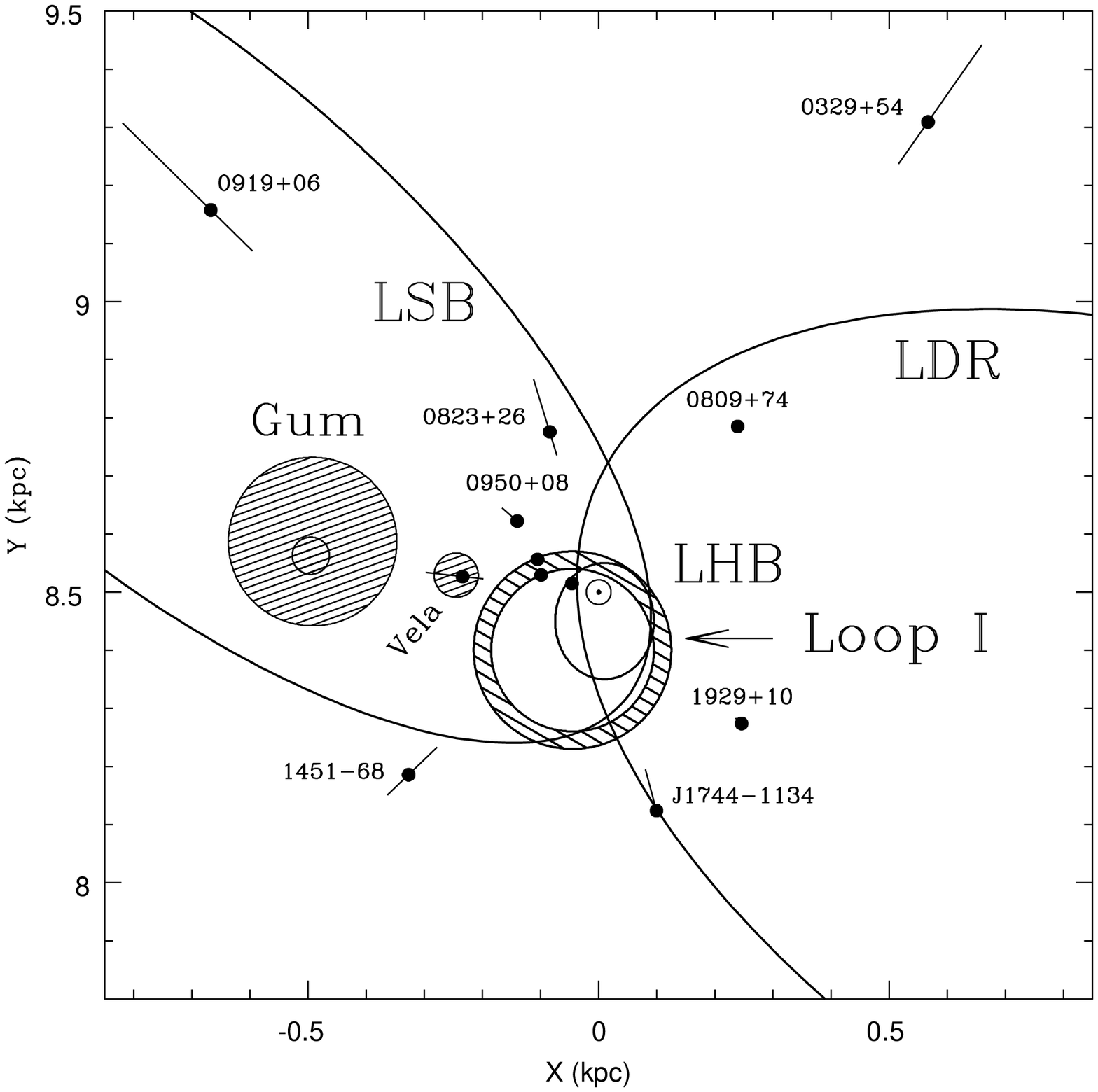}
\figcaption{
\label{fig:lism}
Projection onto the Galactic $x$-$y$ plane of the four local ISM components,
LHB, LSB, LDR and \hbox{Loop~I}.  The filled circles
show the DM-predicted locations of those pulsars
having parallax measurements along with the range allowed from the parallax
measurements.  
The plotted lines that point toward the Sun 
represent the allowed distance ranges from parallax measurements. 
For a few cases, these lines are too short to be visible.
}
\medskip

\epsfxsize=8truecm
\epsfbox{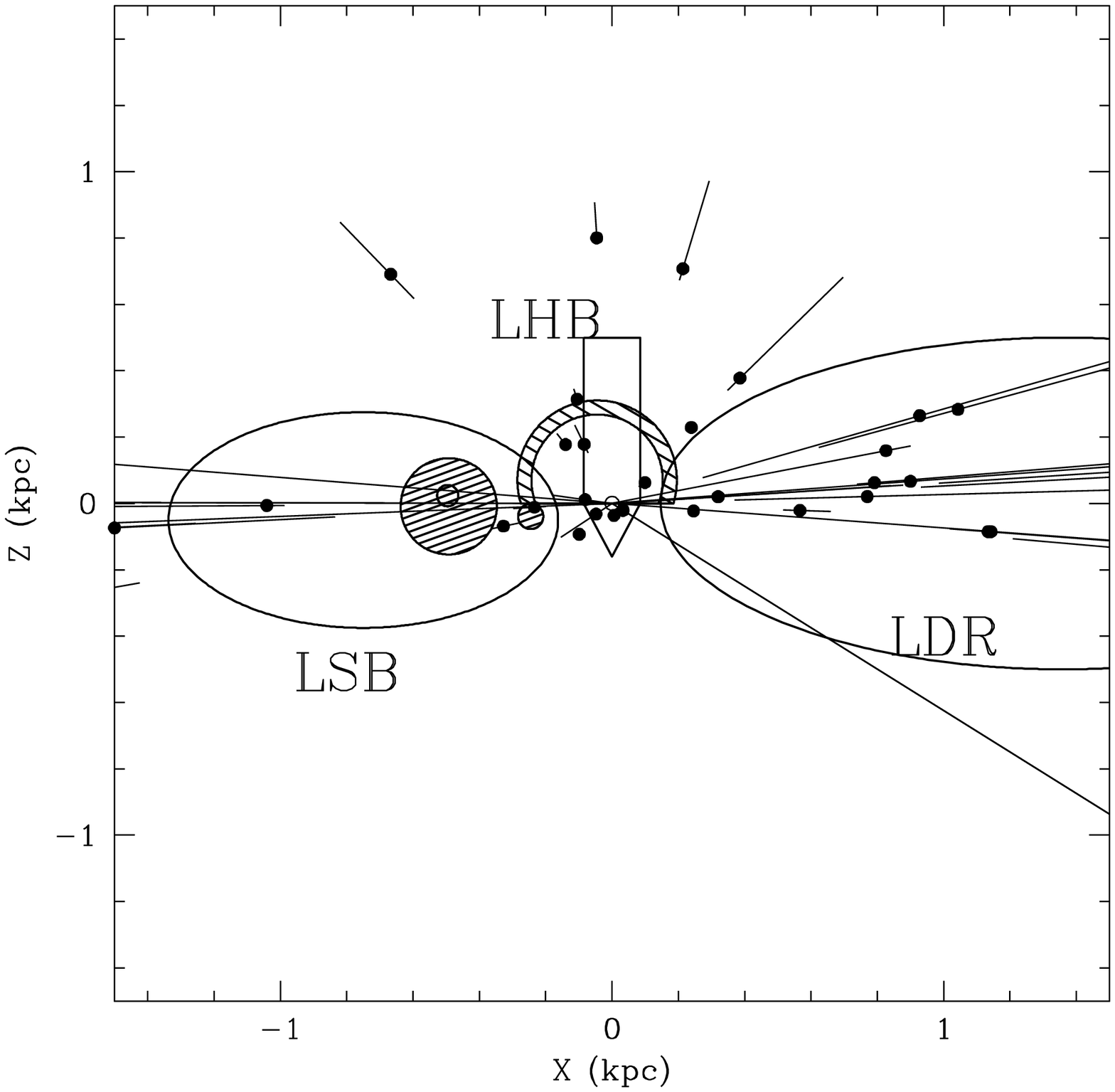}
\figcaption{
\label{fig:lismxz}
Projection onto the $x$-$z$ plane of the four local ISM components,
LHB, LSB, LDR and \hbox{Loop~I}.  
The plotted lines that point toward the Sun 
represent the allowed distance ranges from parallax measurements. 
The points show the predicted locations (using DM) of those pulsars
with parallax measurements.
}

\medskip
\epsfxsize=8truecm
\epsfbox{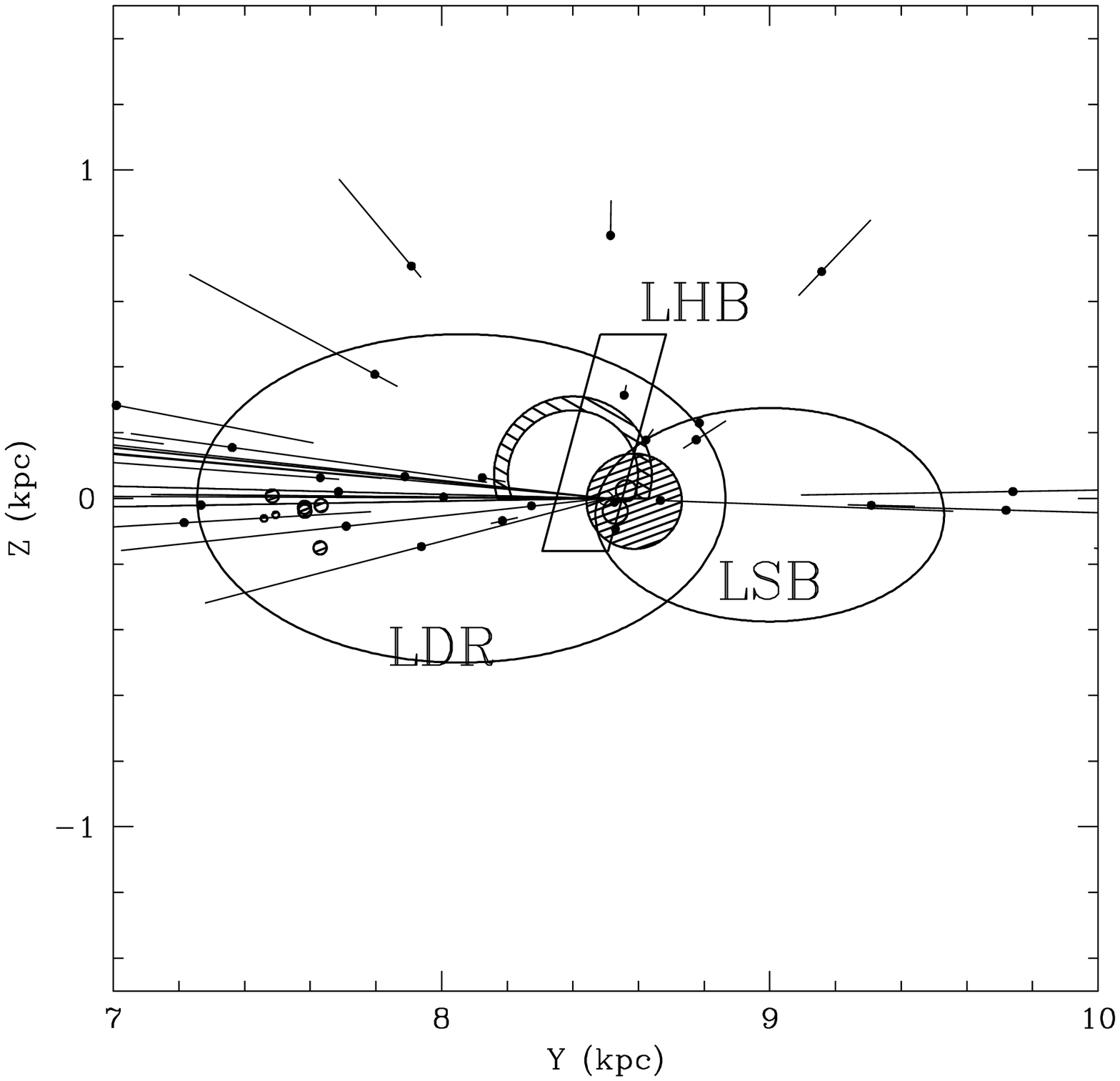}
\figcaption{
\label{fig:lismyz}
Projection onto the $y$-$z$ plane of the four local ISM components,
LHB, LSB, LDR and \hbox{Loop~I}.  
The plotted lines that point toward the Sun 
represent the allowed distance ranges from parallax measurements. 
The points show the predicted locations (using DM) of those pulsars
with parallax measurements.
}

\medskip
\epsfxsize=8truecm
\epsfbox{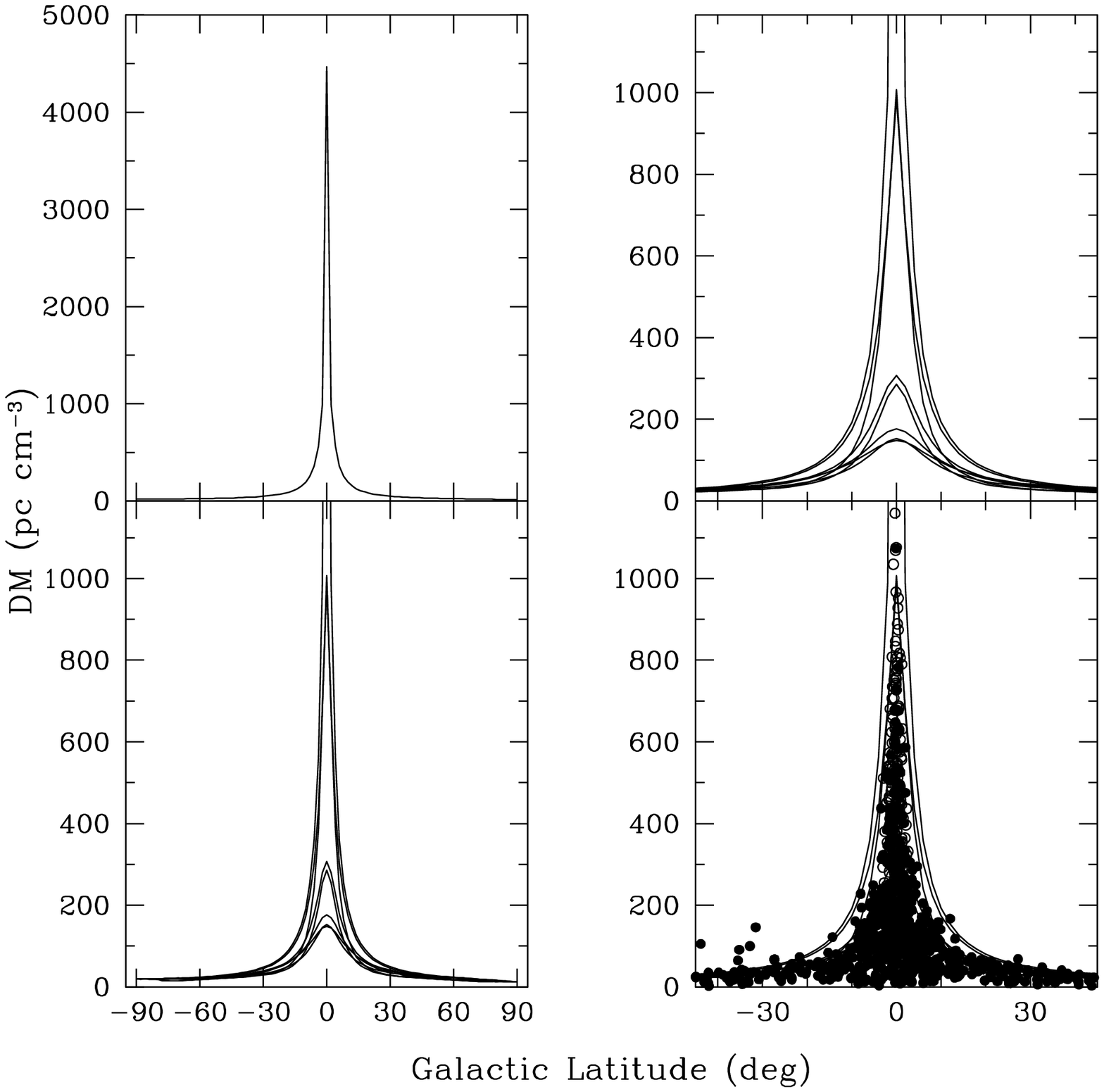}
\figcaption{
\label{fig:dmvsbl.model}
DM plotted against Galactic Latitude for model and data.
Top left:  For Galactic longitude $\ell = 0\arcdeg$ and showing
	the full range of DM. 
Bottom left:  For eight longitudes, 
	$\ell = 0\arcdeg$, $\pm45\arcdeg$, $\pm 90\arcdeg$, $\pm 135\arcdeg$, and~180\arcdeg\ for
	DM $\le 1200$ pc cm$^{-3}$.
Top right: The same eight longitudes for a partial Galactic latitude range.
Bottom right: same as top right but with data points included.
The points about the curves for $b \lesssim -35\arcdeg$ are pulsars in
the LMC and SMC, whose DMs receive contributions from their
host galaxies. 
}

\section{Model Comparisons}\label{sec:comparison}

The final model is considerably more complex than all previously
published electron density models for the Galaxy.  To demonstrate
the contributions of the various model components to the end result
and thus elucidate the need for such components, we compare 
the figures of merit discussed in \S\ref{sec:components} for different
models that exclude one or more components of our final model.
Results in Table~\ref{tab:lddmsm} include log likelihoods and values
for $\nhits$, the number of LOS for which the model yields a distance
within the empirical distance range, $[\dl, \du]$, and
$\nover$, the number of objects for which the model provides 
insufficient electrons to account for \DM.   For each model in
the table, we maximized $\Lambda_d \equiv \log \lpsrd$ by varying
relevant parameters.  The reported value of $\Lambda_d$ was the maximum
value unless a slightly submaximal $\Lambda_d$ yielded a larger
value of $\nhits$  or $\Lambda_s \equiv \log \lpsrs$.  With the model
parameters selected in this way, we then calculated $\nover$.

The models are listed in Table~\ref{tab:lddmsm} in order of 
increasing goodness of fit, taken as a subjective combination
of $\Lambda_d, \Lambda_s, \nhits$ and $\nover$.
As discussed in \S\ref{sec:role},  we consider minimization of
$\nover$ to be a necessary requirement of a good model.   The last
column of Table~\ref{tab:lddmsm} shows a decline from 77 for the
simplest model (a thick, symmetric disk combined with the GC component)
to 6 for the final model.  The 6 objects include 5 in the LMC and SMC, 
which
receive signficant contributions from their host galaxies.
The lone Galactic object (PSR~J1549$+$2110, $\DM = 30$ pc cm$^{-3}$) 
in this group of 6
has only a marginal excess over the model,
$\Delta\DM = 1.3$ pc cm$^{-3}$.   There is a steady increase in
$\nhits$ and decrease in $\Lambda_s$ going down the table.    
Perusal of Table~\ref{tab:lddmsm} also indicates the following.
First, the need for clumps and voids 
is indicated by the sharp drops in $\Lambda_d$ and $\Lambda_s$ in going
from models without these components (Fits~1--3) to those that do
(Fits~4--6).  Inclusion of the LISM components leads to a significant 
increase in $\nhits$ (by design) as do clumps and voids,
spiral arms, and the thin disk.  Spiral arms and the thin disk together       
reduce significantly the scattering likelihood.  Spiral arms produce
a large drop in $\nover$.   There is considerable covariance between
the spiral arms and the thin disk, which are colocated in the inner
Galaxy.  In this regard, the need for a thin disk and the spiral arms
is best demonstrated from $\Lambda_s$ and $\nover$, rather than
from $\Lambda_d$ and $\nhits$.

For comparison, we also show results for TC93 and for the recent model
of GBC01.
For these models, we have fixed
their parameters to the published values.  TC93 performs much less
well than NE2001 in all figures of merit.   GBC01 is an axisymmetric
model constructed using only those pulsars having independent distance
constraints and providing only the local mean electron density.  Therefore
no estimates of scattering may be made with GBC01.   For the three
relevant figures of merit, GBC01 performs less well than either TC93 or
NE2001.

\section{Goodness of Fit}\label{sec:goodness}

Figures~\ref{fig:lism}--\ref{fig:lismyz} show nearby pulsars with
empirical distance ranges (from parallax measurements)
plotted as lines whose extrapolations
intersect the Sun's location.   The filled circles
indicate the distance estimated with NE2001.   As can be seen,
all objects have model-predicted distances
that fall within the allowed parallax distance ranges.
Paper I also discusses model performance.   Figure 8 of Paper I
shows distance estimates and empirical constraints (mostly from
HI absorption) for objects at large distances. 
The model was constructed in part by attempting to get these distances
correct, so the figure  essentially demonstrates self-consistency of
our procedure.  The same can be said for Figure 9 of Paper I,
which shows a scatter plot of estimated distance and empirical
distance range, $[\dl, \du]$.

\section{Individual Lines of Sight}\label{sec:LOS}

In this section we discuss various lines of sight.  We shall elaborate
on the need for a clump or void on some lines of sight, while other
lines of sight provide constraints on the parameters of the clumps or
voids.

We have searched for \ion{H}{2} regions, supernova remnants, and O
stars within~30\arcmin\ of the line of sight to any pulsar requiring a
clump.  Although many \ion{H}{2} regions have angular sizes less than
30\arcmin, they may have extended envelopes that would contribute to a
pulsar's \DM\ (Anantharamaiah~1985) and models of \ion{H}{2} regions
surrounding O stars suggest that they could be larger than 1\arcdeg\
(Miller \& Cox~1993).  Moreover, the size of at least one clump can be
constrained to be at least 15\arcmin\ in size (see the discussion of
PSR~B1849$+$00 below).  Thus, we view 30\arcmin\ as a reasonable
estimate for the size of a clump.


Of course, it is possible that what we have modeled as a single clump
is in fact the combined contribution of many \ion{H}{2} regions or
supernova remnants.  Indeed, many of the pulsars have multiple objects
close to their lines of sight.  Thus, the discussion below should lend
plausibility to our choice to insert clumps along the line of sight to
many pulsars or other sources, though we in many cases cannot
associate a given clump with any particular \ion{H}{2} region or
supernova remnant.  Conversely, in some cases, we believe the case for 
a clump is nearly self-evident, e.g., masers associated with an
\ion{H}{2} region, and we shall not discuss those lines of sight.

We also have searched for \ion{H}{1} shells along the lines of sights
where we have identified voids.  Using 100~pc as a characteristic size
and 3~kpc as a characteristic distance (Heiles~1979), we find that the
typical shell may be able to affect lines of sight within~1--3\arcdeg.

\begin{description}

\item[NGC~6334N and NGC~6334B]
The highly scattered masers in NGC~6334N are probably associated with
the \ion{H}{2} region NGC~6334 while the extragalactic source
NGC~6334B is seen through it.

\item[OH~353.298$-$1.537]
This line of sight does not pass through our Galactic center
component.  We have been able to find only one possible source of
scattering toward the maser, the O star LS~4227, which is at an
estimated distance of~2~kpc (Kilkenny~1993).

\item[OH~359.14$+$1.14]
This line of sight does not pass through our Galactic center
component, though the maser itself is thought to be more distant than
the Galactic center (van~Langevelde \etal~1992).  We have been able to
find only one possible source of scattering toward the maser, the O
star system CCDM~J17381$-$2907AB, which is at an indeterminate
distance.

\item[OH~20.1$-$0.1 OH~40.6$-$0.2, OH~43.80$-$0.13]
These OH masers are associated with \ion{H}{2} regions (Hansen et al.~1993).




\item[PSR~B0138$+$59] The distance constraints from \ion{H}{1} absorption,
$1.9 \le D \le 3.6$ kpc (Graham \etal\ 1974), imply $n_e \approx
0.01$--0.018 cm$^{-3}$, substantially smaller than expected if the
pulsar in fact resides within the Perseus spiral arm.  To allow this
distance and match the scattering, an ad hoc void is needed with
substantial underdensity and smaller $F$ parameter than is typical of
the spiral arms.  We suspect that the pulsar is nearer than the
alleged lower distance estimate of 1.9 kpc.
This low-\DM\ pulsar ($\DM = 35$~pc~cm${}^{-3}$)
has a model distance of~2.2~kpc.  Within~3\arcdeg\ is the
\ion{H}{1} shell GSH~130$+$00$+$15 at an estimated distance of~0.5~kpc
(Heiles~1979).

\item[PSR~B0531$+$21 (= Crab pulsar)]
The Crab pulsar is well-known to be associated with the Crab Nebula.

\item[B0809$+$54] ($\ell = 140.0\arcdeg, b = 31.6\arcdeg$)
The \DM\ to this pulsar implies a LOS average density
of 0.013 cm$^{-3}$ over a (parallax) distance $0.433\pm 0.008$ kpc. 
The scattering is quite low.  To match the distance and \DM,
the model requires a significant path length through the
LDR in our LISM model.   Rickett, Coles \& Markkanen (2000)
used weak scintillation  measurements to constrain the scattering
medium along the LOS, finding consistency either with a uniform,
but weak, Kolmogorov medium or a nonuniform one with a discrete
scattering screen $\sim 0.2$ kpc from the Sun, assuming a distance
(based on TC93) of 0.31 kpc.  The new distance 
(Brisken 2001; Brisken \etal\ 2002)
implies a proportionately larger screen distance and
an even weaker scattering medium ($\log\SM\  = $ kpc m$^{-20/3}$). 
The low  $F$ parameter and low density of the LDR accomodate 
these measurements. 

\item[B0950$+$08] This LOS has the  smallest known scattering measure, owing
to the path length being predominantly through the LHB and LSB,
which have small electron densities and small fluctuation parameters.

\item[PSR~1019$-$5749]
The line of sight to this high-\DM\ pulsar ($\DM >
1000$~pc~cm${}^{-3}$ passes close to a number of \ion{H}{2} regions,
which could individually or in combination account for the need for a
clump along this line of sight.  The \ion{H}{2} region NGC~3199 is
approximately 10\arcmin\ from the line of sight to this pulsar.
Situated in the Carina arm, Deharveng \& Maucherat~(1974) cite a
distance of~3.8~kpc.  Caswell \& Haynes~(1987) also find the \ion{H}{2}
region G283.978$-$0.920 to have a kinematic distance of approximately 
5~kpc.  (This is probably the same \ion{H}{2} region that Wilson et
al.~[1970] name G284.0$-$0.9, for which they find a similar distance.)

%

\item[PSR~J1022$-$5813]
The \ion{H}{2} regions Gum~29, G284.0$-$00.9, and G284.650$-$0.484 are
within 0\fdg5 and at distances of~4--6~kpc (Wilson et al.~1970;
Caswell \& Haynes~1987).  Also close to the line of sight is the
supernova remant SNR~284.3$-$01.8, which Ruiz \& May~(1986) argue is
between~1 and~4~kpc distant and probably more likely to be at the far
end of this distance range.


\item[PSR~B1054$-$62] 
This moderate \DM\ pulsar ($\DM = 323$~pc~cm${}^{-3}$) has a model
distance of approximately 3~kpc.  At approximately the same distance
is the exciting star of the \ion{H}{2} region RCW~55 (Humphreys~1976)
and for the \ion{H}{2} region itself (Brand \& Blitz~1993) Although
there is some possibility for the pulsar to be in front of the
\ion{H}{2} region, the possible biases of the some of the distance
estimates suggest otherwise.  If we do not insert a clump in front of
this pulsar, its distance would be larger than otherwise predicted by
our model.  The photometric distance estimate to the exciting star, if
not corrected properly for dust obscuration, will tend to
underestimate the distance to the \ion{H}{2} region.  Also close to
the line of sight is the UCHII IRAS~10555$-$6242, at a comparable
distance as RCW~55 (Walsh et al.~1997).  The typically small angular
size of UCHIIs, however, makes it unclear whether this particular one
could affect the line of sight to the pulsar.


\item[PSR~B112$-$60]
The line of sight to this pulsar ($\DM = 677$~pc~cm${}^{-3}$) passes
close to a number of \ion{H}{2} regions.  Our model distance for this
pulsar is 15~kpc, and without the addition of a clump, its model
distance would be well outside the Galaxy.  There is a cluster of
\ion{H}{2} regions (G291.205$-$0.267, G291.466$-$0.128, and NGC~3603)
at approximately 8.5~kpc (Caswell \& Haynes~1987).  A second problable
cluster of \ion{H}{2} around~3.5~kpc, related to the Carina arm, also
exists that includes NGC~3576 (Wilson et al.~1970) and the UCHII
IRAS~11097$-$6102 (Walsh et al.~1997).

\item[PSR~J1119$-$6127] Crawford et al.~(2001) argue that this pulsar
is associated with the supernova remnant SNR~292.2$-$00.5.  In
addition the \ion{H}{2} region G291.9$-$0.7 (Wilson et al.~1970;
Caswell \& Haynes~1987) lies close to the line of sight.  Crawford et
al.~(2001) adopt a distance of~5~kpc to the pulsar-SNR association,
though they acknowledge that it is highly uncertain.  The \ion{H}{2}
region is at approximately 10~kpc, and our model estimates a distance
of~16.7~kpc from the \DM\ of the pulsar.  One resolution of these
various distances would be to place multiple clumps in front of the
pulsar, from both the SNR and the \ion{H}{2} region, and locate all of
the objects in the far portion of the Carina spiral arm at a distance
of~8--10~kpc.


\item[PSR~J1128$-$6219]
Our estimated distance to this pulsar ($\DM = 675$~pc~cm${}^{-3}$) is
15.4~kpc.  The following \ion{H}{2} regions are close to the line of
sight and have estimated distances that place them in front of the
pulsar: RCW~60 at 2.4--4~kpc (Humphreys~1974; Caswell \& Haynes~1987)
and G293.027$-$1.031 at~14.5~kpc.  Also close to the line of sight are 
the UCHIIs IRAS~11298$-$6155 at~10.1~kpc and IRAS~11304$-$6206
at~11.1~kpc (Walsh et al.~1997).  Finally, the O stars HD~99897 (in
the \ion{H}{2} region Gum~39) and HD~308570 lie close to the line of
sight, but at indeterminate distances.

\item[PSR~B1131$-$62]
Our estimated distance to this pulsar ($\DM = 568$~pc~cm${}^{-3}$) is
11.8~kpc.  The following UCHIIs are close to the line of sight and
have estimated distances that place them in front of the pulsar:
IRAS~11304$-$6206  at~11.1~kpc and IRAS~11332$-$6258 at~1 or~7.1~kpc
(Walsh et al.~1997).  Also close to the line of sight, but at an
indeterminate distance is the \ion{H}{2} region G294.34$-$1.33.

\item[PSR~J1201$-$6306]
Our estimated distance to this pulsar ($\DM = 683$~pc~cm${}^{-3}$) is
12.7~kpc.  The \ion{H}{2} regions G297.506$-$0.765 at~11.4~kpc and
G297.655$-$0.977 at~11.7~kpc (Caswell \& Haynes~1987) may lie in front 
of this pulsar.

\item[PSR~J1216$-$6223]
Our estimated distance to this pulsar ($\DM = 787$~pc~cm${}^{-3}$)
is 17~kpc, and its estimated distance would locate it outside the
Galaxy if we did not place a clump along the line of sight to it.
Lying close to the line of sight to the pulsar are the \ion{H}{2}
regions G299.016$+$0.148 at an estimated distance of~11.8~kpc and
G298.559$-$0.114 at~11.7~kpc (Caswell \& Haynes~1987) and the UCHII
IRAS~12146$-$6212  at~10.6~kpc (Walsh et al.~1997).  In addition,
three supernova remnants at indeterminate distances also lie close to
this line of sight, SNR~298.6$-$0.0, SNR~299.0$+$0.2, and
SNR~298.6$+$0.0.

\item[PSR~J1305$-$6256]
Our estimated distance to this pulsar ($\DM = 967$~pc~cm${}^{-3}$) is
11.9~kpc.  The SNR~304.6$+$0.1 has an \ion{H}{1} absorption constraint 
on its distance of 9.7~kpc (Caswell et al.~1975), potentially placing
it in front of the pulsar.  Two O stars, at indeterminate distances,
also lie close to the lie of sight, HD~113754 and HD~114122.


\item[B1310$+$18 (in NGC 5024 = M53)]

\item[PSR~B1316$-$60]
At an indeterminate distance, the O star SAO~252232 lies close to the
line of sight to this pulsar.

\item[PSR~J1324-6146]
Close to the line of sight, at an indeterminate distances, are the
\ion{H}{2} region G307.1$+$1.2 and the O star GSC~08995-01924.

\item[PSR~1323$-$62]
Our estimated distance to this pulsar ($\DM = 318.4$~pc~cm${}^{-3}$)
is 5.5~kpc.  Near the line of sight is the O star LSS~3052, which, on
the basis of its $E_{B-V}$, is approximately 3.9~kpc distant
(Turner~1985).  Also close to the line of sight, but at indeterminate
distances, are the O stars GSC~08995-01924 and SS~246.

\item[PSR~B1334$-$61]
Our estimated distance to this pulsar ($\DM = 638$~pc~cm${}^{-3}$) is
9.5~kpc.  The following \ion{H}{2} regions are close to the line of
sight and have estimated distances that place them in front of the
pulsar: G308.647$+$0.579 at~4.4~kpc (or a less likely distance
of~8.1~kpc, Caswell \& Haynes~1987), G308.6$+$0.6 at~4.7 or~7.8~kpc,
and G308.7$+$0.6 at~3.9 or~8.6~kpc (Wilson et al.~1970).  The
supernova remnant SNR 308.7$+$0.0 may also be close enough to the line 
of sight to affect the \DM; Caswell et al.~(1992) adopt a distance
of~6.9~kpc to it.  Finally, close to the line of sight, but at
indeterminate distances are the O star GSC~08995-00021 and the
\ion{H}{2} regions Gum~48c and BBW~27600.


\item[PSR~B1338$-$62]
Kaspi et al.~(1992) argue that this pulsar ($\DM =
730$~pc~cm${}^{-3}$) is associated with the supernova remnant
SNR~308.8$-$0.1, of which SNR~308.7$+$0.0 forms a portion (Caswell et
al.~1992).  The distance, albeit highly uncertain, that they adopt for
the remnant is 6.9~kpc.  Our estimated distance for the pulsar is
11~kpc.  Additional dispersion may be contributed by the \ion{H}{2}
region G309.057$+$0.186, which is at a distance of~3.9~kpc (or a less
likely distance of~8.7~kpc, Caswell \& Haynes~1987).  Further, the O
star GSC~08995-00021 lies close to the line of sight, though at an
indeterminate distance.


\item[J1430$-$6623 \& J1453$-$6413]  The lower bound
$\dl=2.0$ kpc from \ion{H}{1} absorption for J1453$-$6413 (Koribalski \etal\ 1995)
requires a void along the LOS at the Carina-Sagittarius arm.  The 
scattering is overestimated by a factor of $\sim 12$.  The nearby object
J1430$-$6623 requires a clump to bring its distance smaller so as to account
for the scattering (there is no distance constraint on this object).

\item[PSR~B1508$-$57]
Our estimated distance to this pulsar ($\DM = 628.7$~pc~cm${}^{-3}$ is
7.3~kpc.  There are a number of \ion{H}{2} regions close to this line
of sight, but kinematic distance ambiguities make it difficult to
assign reliable distances to them.  Nominally in front of the pulsar
are the \ion{H}{2} regions G321.038$-$0.519 (Caswell \& Haynes~1987)
and G321.0$-$0.5 (Wilson et al.~1970), both at~4.4; however, both
\ion{H}{2} regions could also be approximately 11~kpc distant.
Similarly, nominally behind the pulsar are the \ion{H}{2} regions
G320.379$+$0.139, G320.706$+$0.197, and G320.317$-$0.208 whose
estimated distances are approximately 15~kpc (Caswell \&
Haynes~1987).  Though regarded as less likely, distances between~0.2
and~0.7~kpc are also acceptable for these three \ion{H}{2} regions.
Finally, the \ion{H}{2} region G320.3$-$0.2 and the O star
GSC~08702-00089 are also close to the line of sight, though at
indeterminate distances.

\item[PSR~B1518$-$58]
Close to the line of sight to this pulsar are the supernova remnant
MSC~321.9$-$1.1 and the O star [L64]~240.

\item[B1534$+$12]  The distance to this pulsar is constrained from
the 6-month parallax term in pulse arrival times ($D>0.6$ kpc)
and from  correction for the Shklovsky effect on the orbital  
period derivative of the pulsar (Stairs \etal\ 1998), 
which implies
$D= 1.1\pm 0.2$ kpc.
The lower distance limit $\dl = 0.9$ kpc 
and measured scintillation bandwidth are difficult to reconcile.
With the large scale model, a distance of 0.54 kpc is obtained.
A low density void is needed to reach $\dl$.  But even
with $F=0$ in the void,  the strength of scattering is overestimated
by a factor of 5.   A distance near 0.6 kpc would reconcile this
discrepancy but would then be inconsistent with timing analyses
of the orbit.    We conclude that the line of sight to this pulsar
is likely dominated by conditions interior to a ``blowout'' region.

\item[PSR~J1544$-$5308]
Close to the line of sight to this pulsar are the supernova remnant
MSC~327.4$+$1.0 and the O star [L64]~282.

\item[J1559$-$4438 \& J1600$-$5044]  The minimum \ion{H}{1} absorption distances to
these pulsars are 1.5 kpc and 5.9 kpc, respectively.    The latter
requires a void along the line of sight that, if larger than 200 pc
in transverse extent, also influences J1559$-$4438.   However, the
large implied distance for J1559$-$4438 is inconsistent with the
scintillation bandwidth for electron density components currently in 
the model;  the scattering is overestimated by a factor $\sim 6$.  

\item[J1602$-$5100]  The \ion{H}{1} absorption spectrum to this object
shows a questionable feature at $-$110 km s$^{-1}$ that, if real,
indicates $\dl \approx 7.4\pm0.5$ kpc.   If not real, then
$\dl \approx 5.5\pm0.5$ kpc, a distance consistent with the
present model.

\item[PSR~J1605$-$5257]
Our estimated distance to this pulsar ($\DM = 32$~pc~cm${}^{-3}$) is
1.5~kpc.  There is an UCHII along the line of sight,
IRAS~15596$-$5301, but the estimated distance to the UCHII is either
4.8 or~12.9~kpc, placing it behind the pulsar.  No other objects
appear to be in front of the pulsar that could account for the need
for a clump.

\item[PSR~B1627$-$47]
Close to the line of sight to this pulsar is the supernova remnant
SNR~336.7$+$00.5.


\item[PSR~B1641$-$45]
Our estimated distance to this pulsar ($\DM = 480$~pc~cm${}^{-3}$) is
5~kpc.  There are a number of \ion{H}{2} regions close to the line of
sight of this pulsar that might affect its \DM.  Because of distance
ambiguities, it is not clear that all of these \ion{H}{2} regions can
affect the pulsar's \DM, though.  The following \ion{H}{2} regions
have distances compatible with being in front of the pulsar:
G339.286$+$0.163 at~5.8~kpc, though 13~kpc is also possible;
G339.128$-$0.408 at~3.3~kpc, though 15.4~kpc is also possible if less
likely; G338.921$-$0.089 at~3.6~kpc, though 15.1~kpc is also possible
if less likely (Caswell \& Haynes~1987); and G338.9$-$0.1 at~3.6~kpc,
though 15.1~kpc is also possible (Wilson et al.~1970).  The UCHII
IRAS~16424$-$4531 is at~3.3~kpc, though 15.5~kpc is also a possible
distance for it.  The \ion{H}{2} regions G339.089$-$0.216 at~9.3~kpc
and G339.578$-$0.124 at~16~kpc (though a less likely distance is
2.8~kpc) are unlikely to affect the pulsar's \DM.

Close the line of sight, at an estimated distance of about~1~kpc, is
the highly obscured open star cluster Westerlund~1 (Westerlund~1961; 
Piatti, Bica, \& Clari\'a~1998).  Although this cluster is in front of 
the pulsar and its integrated spectrum contains nebular emission lines 
indicative of significant amounts of ionized gas, it is not clear to
what extent this cluster could affect the \DM\ of the pulsar.  If the
cluster is embedded entirely within higher density gas, the extent of
the cluster's \ion{H}{2} region may be quite small.  Alternately, the
cluster may be on the far side of the obscuring region and producing a 
significant ``blister'' \ion{H}{2} region.

Also close to the line of sight, though of indeterminate distances,
are the supernova remnant SNR~339.2$-$0.4 and the \ion{H}{2} region
G339.6$-$0.1.

The O star HD~151018 also lies close to the line of sight to this
pulsar, at an approximate distance of~1.6~kpc, and Miller \&
Cox~(1993) predict the radius of its \ion{H}{2} region of order 100~pc
($= 3\fdg6$).


\item[PSR~B1643$-$43]
This pulsar is associated with the supernova remnant SNR~341.2$+$0.9
(Frail, Goss, \& Whiteoak~1994; Giacani et al.~2001).


\item[PSR~B1703$-$40]
Our estimated distance to this pulsar ($\DM = 360$~pc~cm${}^{-3}$) is
4.4~kpc.  The following \ion{H}{2} regions may lie in front of this
pulsar, though because of distance ambiguities, these regions may also
lie behind the pulsar: G345.450$+$0.209 at~1.7~kpc, though 17.7~kpc is
also a possible distance; G345.645$+$0.010 at~1.3~kpc, though 18.1~kpc
is also possible; and G346.109$-$0.028 at~1.5~kpc, though 17.9~kpc is
also possible (Caswell \& Haynes~1987).  In addition the UCHII
IRAS~17031$-$4037 is at~1.5~kpc, though 18~kpc is also a possible
distance (Walsh et al.~1998).  The \ion{H}{2} regions G345.555$-$0.042
and G345.827$+$0.041 have distances estimates of approximately 18~kpc,
though distances around~1~kpc are also allowed (Caswell \&
Haynes~1987).

Also close to the line of sight, though of indeterminate distances,
are the O star GSC~07873-00112 and the supernova remnant
SNR~345.7$-$0.2.

\item[J1709$-$4428]  The \ion{H}{1} absorption lower-distance bound
of 1.8 kpc (Koribalski \etal\ 1995) requires a void within
the Carina-Sagittarius arm;  the model still overestimates
the scattering by a large factor ($\sim 70$). 

\item[PSR~B1715$-$40]
Close to the line of sight to this pulsar, though of indeterminate
distances, are the O stars GSC~07874-00170 and  GSC~07874-01001.

\item[PSR~B1718$-$36]
Our estimated distance to this pulsar ($\DM = 416$~pc~cm${}^{-3}$) is
4.3~kpc.  Potentially in front of this pulsar is the \ion{H}{2} region 
G350.813$-$0.019 at~1~kpc, though distance ambiguities mean that it
could also be at~18.7~kpc (Caswell \& Haynes~1987).  Also close to the 
line of sight, but at indeterminate distances, are the \ion{H}{2}
region RCW~128, the O star HD~319734, and the supernova remnant
SNR~351.2$+$0.1 (which itself has a central point source and may
contain another pulsar, Becker \& Helfand~1988).


\item[PSR~B1714$-$34]
Close to the line of sight to this pulsar, though of indeterminate
distances, are the \ion{H}{2} regions RCW~130 and Sh~2-10 and the O
double star \hbox{IDS~17118$-$3418B}.

\item[PSR~B1727$-$33]
Our estimated distance to this pulsar ($\DM = 256.5$~pc~cm${}^{-3}$)
is 3.5~kpc.  We have been unable to identify any ionizing source that
is definitively in front of this pulsar.  The \ion{H}{2} region
G354.200$-$0.054 is behind the pulsar, with ambigious distance
estimates of~6.5 or~13.4~kpc, as is G354.486$+$0.085 with a large,
though uncertain distance estimate (Caswell \& Haynes~1987).  The
UCHII IRAS~17279$-$3350 is close to the line of sight, but of
indeterminate distance.

\item[PSR~B1736$-$31]
Our estimated distance to this pulsar ($\DM = 600$~pc~cm${}^{-3}$) is
6.2~kpc.  We have been unable to identify any ionizing source that is
definitively in front of this pulsar, though, the UCHII
IRAS~17352$-$3153 is at a distance of either 8.9 or~11.3~kpc (Walsh et
al.~1997).


\item[PSR~J1745$-$3040]
Our estimated distance to this pulsar ($\DM = 88$~pc~cm${}^{-3}$) is
1.9~kpc.  The O star HD~316256 at~0.92~kpc (Kilkenny~1993) is close to
the line of sight.  Also close to the line of sight, though of
indeterminate distances, are the \ion{H}{2} region G358.664$-$0.575
and the supernova remnant SNR~359.0$-$0.9 (though the latter is
usually taken to be in or near the Galactic center).
  

\item[PSR~B1746$-$30]
Our estimated distance to this pulsar ($\DM = 510$~pc~cm${}^{-3}$) is
5.7~kpc.  Along the line of sight is the O emission star HD~316341
at~1.9~kpc (Kozok~1985).  Also close to the line of sight, though of
indeterminate distances, are the \ion{H}{2} regions G359.158$-$0.881
and G359.281$-$0.826 and the O star LS~4424.

\item[J1757$-$2421 = B1757$-$24 (The Duck Pulsar)]  
The lower distance bound of 4.3 kpc, from \ion{H}{1} absorption against the 
synchrotron nebula surrounding the pulsar (Frail \etal\ 1994b) 
requires an underdensity
along this LOS in the spiral arms intersected
(Carina-Sagittarius and Crux-Scutum), but with sizable fluctuation
parameter ($F\sim 10$) to match the scattering.
Noncircular motions in the \ion{H}{1} gas could alter this distance
bound given the small Galactic longitude ($\ell = 5.3\arcdeg$).

\item[J1759$-$2205]  The DM of this pulsar is nearly identical to that
of J1757$-$2421 while  the pulse broadening time is smaller by a factor
of four.
The scattering requires a smaller F parameter along this line of
sight through the Carina-Sagittarius spiral arm.

\item[PSR~J1807$-$2715]
Close to the line of sight to this pulsar, though of an indeterminate
distance, is the supernova remnant SNR~004.2$-$3.5.

\item[PSR~B1758$-$23]
This pulsar has been associated with the supernova remnant W28 (Frail,
Kulkarni, \& Vasisht~1993; Kaspi et al.~1993).  In addition the O star
HD~164492B is close to the line of sight and at a distance of~1.4~kpc
(Lindroos~1985).  There are a few \ion{H}{2} regions whose distances
place them nominally beyond the pulsar, though the uncertainties in
deriving kinematic distances toward the inner Galaxy may make these
distance estimates highly uncertain: RCW~145 with an estimated
distance of~2.7--13~kpc (Vilas-Boas \& Abraham~2000) and
G6.528$+$0.076, G6.6$-$0.1, and G6.7$-$0.2 all of which have distance
estimates around 3.5~kpc (Wilson et al.~1970).  There are also
numerous \ion{H}{2} regions with indeterminate distances close to the
line of sight: Sh~30, G6.553$-$00.095, G6.565$-$00.297,
G6.667$-$00.247, G6.979$-$00.250, G7.299$-$00.116, G7.002$-$0.015,
G7.236$+$0.144, RCW~147, Ced~151, and Gum~74a, and two O stars
LS~4538 and HD~313599.



\item[PSR~B1815$-$14]
Our estimated distance to this pulsar ($\DM = 625$~pc~cm${}^{-3}$) is
6.9~kpc.  Close to the line of sight is the star cluster NGC~6611,
which contains a number of~O and~B stars, at a distance of~1.6~kpc
(Kamp~1974).  Also close to the line of sight, although at an
indeterminate distance, is the \ion{H}{2} region G016.463$+$0.966.


\item[PSR~1820-14]
Our estimated distance to this pulsar ($\DM = 648$~pc~cm${}^{-3}$) is
4.7~kpc.  Close to the line of sight is the O star LS~4971 at an
estimated distance of~1.3~kpc (Kilkenny~1993).  Also close to the line 
of sight, though of indeterminate distances, are the \ion{H}{2}
regions G16.893$+$0.126, G16.941$-$0.073, and G17.250$-$0.195.

\item[PSR~B1830$-$08]
This pulsar may be associated with the supernova remnant W41 (Clifton
\& Lyne~1986).  Even if not associated with the SNR, there are
UCHIIs that may be in front of it, modulo distance ambiguities:
IRAS~18317$-$0845, IRAS~18317$-$0845, and IRAS~18324$-$0820 all have
distance estimates between~5 and~6~kpc, though they could also be
between~12 and~13~kpc distant (Walsh et al.~1997).  In addition, the
UCHIIs IRAS~18310$-$0825 and IRAS~18310$-$0825 have distance estimates 
of~6.4~kpc, placing them nominally just beyond the pulsar, though they 
may also be 12~kpc distant (Walsh et al.~1997).  Finally, the
\ion{H}{2} regions G23.281$+$0.298, G22.935$-$0.072, G23.000$+$0.219,
G23.162$+$0.023, G23.072$-$0.248, G23.254$-$0.268, G23.42$-$0.21,
G23.538$-$0.041, G23.613$-$0.376, G23.706$-$0.202, and G23.817$+$0.224 
and the supernova remnant SNR~23.6$+$0.3 are close to the line of
sight but with indeterminate distances.
Also consistent with being in front of this pulsar is the \ion{H}{2}
region G23.437$-$0.207 at a distance of~6.6 $\pm$ 0.5~kpc (Kolpak et
al.~2003).



\item[PSR~B1832$-$06]
While close to the SNR~24.7$+$0.6, this pulsar is probably not
associated with it (Gaensler \& Johnston~1995).  In addition there are
numerous \ion{H}{2} regions, though of indeterminate distances, close
to the line of sight of this pulsar:  RCW~173, BBW~35201, Sh~60,
G25.294$+$0.307, G24.91$+$0.57, G24.905$+$0.431, G25.293$+$0.306, and
G25.161$+$0.069.


\item[PSR~J1823$-$0154]
Close to the line of sight to this pulsar is the \ion{H}{2} region
(UCHII?) IRAS~IRAS 18216$-$0156.

\item[PSR~B1839$-$04]
There are a number of \ion{H}{2} regions and supernova remnants close
to this line of sight, albeit of indeterminate distances.  These
include G27.975$+$0.080, G27.997$+$0.317, G28.001$-$0.031,
G28.146$+$0.146, G28.312$-$0.023, G28.400$+$0.478, G28.440$-$0.002,
G28.600$+$0.015, G28.638$+$0.194, G28.658$+$0.030, G28.787$+$0.252,
G28.801$+$0.174, and IRAS~18408$-$0348 (an UCHII?), which are
\ion{H}{2} regions, and G28.600$-$0.126 and AX~J1843.8$-$0352, which
are SNRs.

\item[PSR~B1845$-$01]
Our estimated distance for this pulsar ($\DM = 159$~pc~cm${}^{-3}$) is
3.8~kpc.  Although there are a number of objects close to its line of
sight, few of the objects with estimated distances are in front of the
pulsar.  One object potentially in front of the pulsar is
IRAS~18452$-$0141 whose distance is estimated to be 1.2~kpc, though
distance ambiguities mean that it might also be at~15.6~kpc (Walsh et
al.~1998).  Other \ion{H}{2} regions close to the line of sight, but
at indeterminate distances, are G30.9$+$0.1, G31.130$+$0.284,
G31.131$+$0.284, G31.1$-$0.1, G31.2$-$0.1b, G31.401$-$0.259,
G31.411$+$0.309, G31.602$-$0.227, G31.607$+$0.334, and the UCHIIs
IRAS~18456$-$0129 and IRAS~18469$-$0132.  (\ion{H}{2} regions close to
the line of sight, but nominally behind the pulsar include G31.0$-$0.0
at~8~kpc, G31.0$+$0.1 at~6.6~kpc, G31.2$-$0.1 at~12~kpc [Kuchar \&
Bania~1994], G31.28+00.06 at~8.5~kpc [Caswell \& Haynes~1983],
G31.3$+$0.1 at~7.3~kpc, G31.412$+$0.308 at~7.2~kpc [Kolpak et
al.~2003], G31.6$+$0.1 at~6.5~kpc [Kuchar \& Bania~1994], and the
UCHIIs IRAS~18446-0150 at~7.6 or 9.1~kpc, IRAS~18449-0115 at~8.5~kpc,
and IRAS~18456-0129 at~8.5~kpc [Walsh et al.~1998].)


\item[PSR~B1849$+$00 and extragalactic source B1849$+$00]
Our estimated distance to this pulsar ($\DM = 680$~pc~cm${}^{-3}$) is
8.4~kpc.  Close to this line of sight are the \ion{H}{2} regions
G33.12$-$0.08 and G33.2$-$0.0, both at~7.1~kpc, (Kuchar \& Bania~1994)
and the supernova remnant SNR~33.7$+$0.0, which is at a distance of at
least 7~kpc (Caswell et al.~1975).  Also close to the line of sight,
though at indeterminate distances, are the \ion{H}{2} region
G33.498$+$0.196, the UCHII IRAS~18504$+$0025, and the supernova
remnant SNR~33.1$-$0.1 and the \ion{H}{2} region G32.80$+$0.19 which
has an upper distance limit of~15.6~kpc (Wink, Altenhoff, \&
Mezger~1982).  (Also close to the line of sight, though nominally
behind the pulsar, are the \ion{H}{2} region G33.4$-$0.0 at~9.5~kpc
[Kuchar \& Bania~1994] and the supernova remnant SNR~33.6$+$0.1
at~10~kpc [Frail \& Clifton~1989].)

The line of sight to this pulsar also provides some constraints on the 
size of a (single) clump.  Located 13\arcmin\ away is the
extragalactic source B1849$+$005, which is one of the most heavily
scattered sources known (Spangler \& Cordes~1988).  Thus, to the
extent that the enhanced dispersion and scattering along this line of
sight can be modeled as due to a single clump, the clump must be at
least 13\arcmin\ in diameter, equivalent to a linear diameter of~15~pc 
at~4~kpc (halfway to the pulsar).


\item[PSR~B1859$+$03]  The scattering requires a smaller $F$ parameter
along a substantial fraction of the LOS to this pulsar, which has
a lower distance bound of 6.8 kpc.
The model estimated distance to this pulsar
($\DM = 403$~pc~cm${}^{-3}$) is 7.3~kpc.  Within~3\arcdeg\ is the
\ion{H}{1} shell GSH~036$+$01$-$21 at an estimated distance of~0.5~kpc
(Heiles~1979).

\item[OH35.2$-$1.7 \& PSR~B1900$+$01]  A modest constraint on the
angular scale of a clump can be set from these two lines of sight.
The OH maser requires a clump along the LOS while the pulsar, $\sim
0.6\arcdeg$ away, does not.

\item[PSR~B1900+01 and OH~35.2$-$1.7]
Our estimated distance to this pulsar ($\DM = 246$~pc~cm${}^{-3}$) is
3.3~kpc.  At nearly the same distance is the \ion{H}{2} region W~48
(Radhakrishnan et al.~1972), which given distance uncertainties we
regard as potentially capable of perturbing the pulsar's \DM.  Also
close to the line of sight, though at indeterminate distances, are the
\ion{H}{2} region G35.346$-$01.827 and the supernova remnant
SNR~35.0$-$1.1.  The OH maser is itself associated with an \ion{H}{2}
region (Hansen et al.~1993), but probably in front of W~48 (Caswell \& 
Haynes~1983).

\item[PSR~B1907$+$10]
Close to the line of sight for this pulsar is the \ion{H}{2} region G45.200$+$0.740.

\item[PSR~B1919$+$21]
Close to the line of sight for this pulsar is the supernova remnant SNR~55.7$+$3.4.

\item[PSR~B1929$+$20]
Close to the line of sight for this pulsar is the supernova remnant
SNR~55.6$+$0.7.

\item[B1930$+$22]  The \ion{H}{1}-absorption distance constraints
of 9.8 - 14.4 kpc are based on low signal to noise features
at negative velocity.   Frail \etal\ (1991) state that the
-16 km s$^{-1}$ feature that corresponds to 9.8 kpc in
a circula-Galactic rotation model
could be a local, high-velocity cloud, in which
case the pulsar could be nearer than 9.8 kpc.  NE2001 places
the pulsar at  $D = 7.4$ kpc. 

\item[PSR~B1951$+$32]
This pulsar is associated with the supernova remnant CTB~80 (Clifton
et al.~1987).

\item[PSR~B1952$+$29]
Close to the line of sight for this pulsar is the supernova remnant SNR~65.7$+$1.2.

\item[PSR~2053$+$36]
Close to the line of sight for this pulsar is the \ion{H}{2} region G78.81$-$5.71.

\item[B2127$+$11A-H (M15)]  The DMs to pulsars in the globular
cluster  M15 are larger
by $\sim 13$ DM units than the smooth model can accomodate
for $b = -27.3\arcdeg$.   We have added a clump at 2 kpc distance,
about that of the Cygnus superbubble, but well out of the Galactic
plane.

\item[B2210$+$29]
A \DM\ excess $\Delta\DM \approx 10$ pc cm$^{-3}$ persists
while varying relevant model components through credible ranges.
We have therefore included a clump along the line of sight at a distance
of 2 kpc, about that of the Cygnus superbubble, but well out of
the Galactic plane at $b = -21.7\arcdeg$.

\item[OH12.2$-$0.1]  The scattering to this OH maser is overestimated
by a factor of three using its attributed distance of 16.1 kpc
(Hansen \etal\ 1993).    We question the accuracy of this distance 
estimate.

\item[Objects seen through the Cygnus region]
These objects include Cyg~X-3 and several extragalactic sources.  The
Cygnus region is thought to be the tangent to a spiral arm (Bochkarev
\& Sitnik~1985) and Spangler \& Cordes~(1998) have shown how the
Cygnus OB1 association is probably responsible for a large fraction of
the scattering along these lines of sight.


\item[Objects in the Galactic Center]  The GC scattering volume
described earlier adequately describes the angular broadening
of Sgr A*, 
the Galactic Center transient source (Zhao \etal 1991), 
and many of the OH masers (van Langevelde \etal\ 1992; Frail \etal\ 1994).
However,   the angular broadening for some of the masers is underestimated
by a factor of a few.   The scatter in the angular broadening predictions
undoubtedly derives, in part, from their unknown true distances.   We have 
uniformly attributed a distance of 8.5 kpc to all GC sources.  
The GC scattering region $\sim 0.1$ kpc in radius, so changes in true
distance of this order, which are expected for the population
of OH/IR stars in the GC,  will produce significant changes in the amount
of scattering at the level seen.

By contrast, G359.87$+$0.18 (Lazio \etal\ 1999), thought to be 
an extragalactic source, shows much less scattering, 
$\theta_d\sim 2$ arc sec at 1.0 GHz, compared to an expected 
$\sim 300$ arc sec.    This indicates that the GC scattering volume
is either patchy or does not cover the LOS to this source.  We adopt
the latter interpretation and use it to attribute an offset of
the scattering volume of 20 pc in the $-z$ direction.  We also use
a scale height of 26 pc. 

\item[Objects in the LMC and SMC]   Five objects in our sample
reside in the LMC or SMC,  as is consistent with their DMs being
too large to be accounted for by our final model; augmentation
of the particular lines of sight with \ion{H}{2} clumps is unwarranted because
the lines of sight are well out of the Galactic plane.   
We estimate the \DM\ excess attributable
to their host galaxies, obtained
by subtracting the Galactic contribution
(with \DM\ and $\Delta\DM$ in standard units of pc cm$^{-3}$):
J0045$-$7319, $\DM = 105, \Delta\DM = 62$;
B0456$-$69, $\DM =  91, \Delta\DM = 41$; 
B0502$-$66, $\DM =  65, \Delta\DM = 16$;
B0529$-$66, $\DM = 100, \Delta\DM = 48$;
and
B0540$-$69, $\DM = 146, \Delta\DM = 91$.

\end{description}

\section{Discussion of the Model \& Its Limitations}\label{sec:model.discussion}

The NE2001 model represents a clear improvement upon the TC93 model.
First, the distance estimates obtained from the model agree with
available distance constraints for nearly all pulsars with such
constraints (Figures~9 and~10, Paper~I).  Second, none of the
parameters of the large-scale components are indeterminate (e.g., as
was the case with the thick disk for TC93).  The cost of these
improvements has been an increase in the complexity of the model,
particularly with respect to the number and location of clumps and
voids.  We believe, however, that this additional complexity is
motivated both by the quantity of data and astrophysically
(\S\ref{sec:LOS}).  As such our model embodies the long-known fact
that a small number of pulsars or extragalactic sources have
anomalously large DMs or scattering properties or both due to
intervening \ion{H}{2} regions or supernova remnants.

The typical distance uncertainty, e.g., from unmodeled clumps, is
perhaps slightly less than 20\% (Figure~12 of Paper~I).  This is
slightly better than the 25\% uncertainty estimated by TC93 for their
model.  Nonetheless, given the available data we consider it unlikely
that future models will be able to improve much upon this typical
distance uncertainty, without a substantial change in the nature of
the available data.  The DM-independent distance constraints that we
use to calibrate the model are dominated by \ion{H}{1} absorption
measurements and associations (with supernova remnants or globular
clusters).  The typical accuracy of these constraints is certainly no
better than 10\% and sometimes can be much worse.  Only 10\% of the
distance constraints come from pulsars with parallaxes, either
interferometric or timing, which can have accuracies of~5\% or better.
Until the number of parallaxes becomes comparable to or dominates the
\ion{H}{1} absorption measurements, substantial improvement in the
distance accuracy is unlikely from NE2001 and future models.

The introduction, location, and parameters of the clumps and voids
remains ad hoc to some degree.  On the one hand, this would appear to
limit the predictive power of the model.  For instance, should a
newly-discovered pulsar be considered to be affected by a clump or
void in front of a known pulsar that is nearby in angle?
We regard this as an
opportunity to improve the model.  Discovery of new pulsars will
assist in constraining the location and sizes (and numbers) of clumps
and voids.

Finally, in our model we have assumed that the Sun is in the Galactic
plane and located at the conventional distance of 8.5~kpc from the
Galactic center.  In contrast, current estimates are that the Sun is
approximately 20~pc above the Galactic plane \cite{hl95} and is
located between~7 and~8~kpc from the Galactic center (Reid 1993;
Olling \& Merrefield 1998).  Discerning these effects in the current
data may be difficult due to various selection effects in the
different pulsar surveys, but we anticipate that future editions of
the model will attempt to incorporate these offsets in the solar
location.

Nonetheless, we can estimate crudely how these offsets would affect
the model.  The DM vertical to the Galactic plane is approximately
20~pc~cm${}^{-3}$ (Figure~\ref{fig:dmz}), which is dominated by the
thick disk component.  Its scale height is 0.95~kpc.  Thus we would
expect an approximately 2\% difference ($\approx 0.4$~pc~cm${}^{-3}$)
in the DMs for northern and southern hemisphere pulsars.  Similarly,
decreasing the radial scale lengths of all of the components by 6\%
could be expected to increase the nominal electron densities by a
similar factor.

\section{Discussion of Future Approaches}\label{sec:discussion}

While we consider NE2001 to be an improvement over TC93, we foresee a
number of probable developments that will allow NE2001 to be improved
further.  We group these improvements into increases in the quantity
of data and improvements in the modeling.  Perhaps most important
increase in the quantity of data will be an increase in the number of
pulsar parallaxes, both from timing pulsars in the Parkes multibeam
sample and from large interferometric programs.  DM-independent
distances provide crucial calibration information for NE2001 or any
successors, and we regard it as likely that the number of pulsars with
DM-independent distances will double in the next few years.  We have
made use of only the positions and DMs of pulsars discovered in the
Parkes multibeam sample.  Efforts are underway to measure the
scattering along the lines of sight to many of these pulsars, which
could increase the number of lines of sight with measured SMs by
roughly 50\% or more.  The advent of the Green Bank Telescope and the
refurbished Arecibo telescope suggest the possibility of conducting a
northern hemisphere equivalent of the Parkes multibeam sample, which
could increase the number of pulsars by at least another 50\%.

Although we have provided a
formalism for comparing DM and SM to EM, we have made little use of
it.  Future work to include observational constraints on EM, e.g.,
from H$\alpha$ surveys, has the potential of producing a better model
at least locally.

Finally, Figure~11 of Paper~I suggests that the large-scale structure
of the Galaxy may be able to be determined \textit{ab inito}, provided
that a sufficient number of lines of sight exist.  Rather than
imposing a large-scale structure as done both here and previously, the
presence and location of large-scale components, particularly the
spiral arms, could be determined.  Future pulsar surveys may approach
the number of lines of sight required to employ this approach.

\acknowledgments

We thank 
Z. Arzoumanian, R. Bhat, F.~Camilo, S. Chatterjee,
D. Chernoff, Y. Gupta, S. Johnston, F. J. Lockman, R. N. Manchester, and
B. Rickett 
for useful discussions.  Our research is
supported by NSF grants AST 9819931 and 0206036
and by the National Astronomy and
Ionosphere Center, which operates the Arecibo Observatory under
cooperative agreement with the \hbox{NSF}.  Basic research in radio
astronomy at the NRL is supported by the Office of Naval Research.

\bigskip


\begin{deluxetable}{cccccccccrrcc}
\footnotesize
\tablecolumns{13}                
\tablewidth{0pc}  
\tablecaption{ Likelihood Fits to Distance, \DM\ \& Scattering  Data 
\label{tab:lddmsm}}
\tablehead{
\\
                  &\multicolumn{7}{c}{Components Included} &
                   \multicolumn{4}{c}{Figures of Merit} \\
\\
     \cline{2-8}  \cline{10-13} \\
\colhead{$\rm Model$} & 
	\colhead{Thick} & 
	\colhead{Thin} & 
	\colhead{LISM} & 
	\colhead{Arms} & 
	\colhead{GC} & 
        \colhead{Clumps} & 
        \colhead{Voids} && 
	\colhead{$\Lambda_d$} & 
	\colhead{$\Lambda_s$} & 
        \colhead{$\nhits^{\tablenotemark{d}}$} & 
        \colhead{$\nover^{\tablenotemark{e}}$} \\ 
	&\colhead{Disk} & 
	\colhead{Disk} \\
\\
}
\startdata
\\
NE2001$^{\tablenotemark{a}}$  \\
0 &  --- & \ok  & \ok & \ok & \ok & \ok & \ok &  & 
	 $-3079$ & $-3162$ & 68  & 380\\
\\
1 &  \ok & ---  & --- & --- & \ok & --- & --- &  & 
	$-880$ & $-5814$ & 59  & 77\\ 
\\
2 &  \ok & \ok & --- & --- & \ok & --- & --- &  & 
	$-824$ & $-3512$ & 60 & 70    \\ 
\\
\enddata
\tablenotetext{a}
{Fits were made while including subsets of the components included in 
the complete model.
 LISM and GC parameter values were held fixed for all fits.  
Results are reported for the 
best fit defined as a solution that maximizes the log
likelihood  for distance and \DM\ data,
$\Lambda_d = \log \lpsrd$, while also maximizing $\nhits$.
The accompanying total log likelihood for scattering data,
$\Lambda_s = \log \ls$ is maximized within the subset of solutions
that maximizes $\Lambda_d$.   
}
\tablenotetext{b}
{TC93 = the model of Taylor \& Cordes 1993, having two axisymmetric
components and spiral arms. The 13 parameters were held fixed to their 
published values.}
\tablenotetext{c}
{GBC01 = axisymmetric model of G{\'o}mez, Benjamin \& Cox 2001.
The 6 parameters were held fixed to their published values. This model
yields estimates only for the local mean electron density, so scattering
data cannot be modeled and $\Lambda_s$ is undefined.}
\tablenotetext{d}
{$\nhits$ is the number of objects for which the model provides
a distance estimate within the empirical range, $[\dl, \du]$.
The maximum possible is 119 (c.f. Table 1 of Paper I).}
\tablenotetext{e}
{$\nover$ is the number of objects out of 1134 (c.f. Table 1 of Paper I) 
having DM estimates
for which the measured DM exceeds that of the model when integrating
to effectively infinite distance.}
\end{deluxetable}

\begin{deluxetable}{lrrrrrr}
\tabletypesize{\footnotesize}
\tablecolumns{7}
\tablewidth{0pc}
\tablecaption{Clump Parameters for Pulsar Lines of Sight \label{tab:clumps}}
\tablehead{
\\
\multicolumn{1}{l}{ LOS} & 
\multicolumn{1}{c}{$\ell$} & 
\multicolumn{1}{c}{$b$} &  
\multicolumn{1}{c}{$\dc$} & 
\multicolumn{1}{c}{$\DM_c$} &  
\multicolumn{1}{c}{$\log\SM_c$} & 
\multicolumn{1}{c}{$r_c$}  \\
&
\multicolumn{1}{c}{ (deg)} & 
\multicolumn{1}{c}{(deg)} & 
\multicolumn{1}{c}{(kpc)} & 
\multicolumn{1}{c}{(pc cm$^{-3}$)} & 
\multicolumn{1}{c}{(kpc m$^{-20/3})$} & 
\multicolumn{1}{c}{(kpc)}  \\ 
\\
}
\startdata
\input tab.neclumpN.P1.txt
\enddata
\end{deluxetable}

\input iss_table_psr.tex

\input iss_table_gal.tex

\input iss_table_xgal.tex


\clearpage

\appendix 
\setcounter{table}{0}

\section{Scattering Measure Definition and Estimation}\label{app:sm}

Here we derive our expressions for the scattering measure and its
relationship to observable quantities. 
To have a consistent estimate of scattering from angular broadening, 
temporal broadening, and scintillation measurements, we account
for the fundamental differences in the kinds of measurements.  VLBI of
extragalactic sources requires consideration of plane waves impinging on a
Galactic scattering medium while spherical waves must be considered for  
pulsars that are embedded in the medium.  We define a scattering measure as the
integral over path length $D$,
\be
SM = \int_0^D ds \,\cnsq(s),
\ee
where $\cnsq$ is the coefficient in the wavenumber spectrum.   

\be
P_{\delta n_e} (q) = C^2_n q^{-\beta} \exp [-(q\ell_1 /2)^2],\qquad q_0 \le
q < \infty. 
\ee
The spectrum is a power law with index $\beta =11/3$ for a Kolmogorov spectrum
and incorporates an `inner scale', $\ell_1$ and a lower wavenumber cutoff 
$q_0=2\pi/\ell_0$, where $\ell_0$ is the `outer scale'. 
Here we derive estimators for $SM$ in terms of observable quantities.

\subsection{A1. Angular Broadening}

Consider the phase structure functions implied by eqn (A2).  For a plane wave
impingent on a medium with scattering measure $SM$ we have (Coles {\it et al.}
1987) for $2 < \beta< 4$ and assuming $q_0\,\,b\ll 1$ for all $b$ of interest

\be
D_p (b) = 8\pi^2 r_e^2 \lambda^2 SM \times \cases{
g(\beta) b^2\ell_1^{\beta-4} &$b\ll\ell_1$\cr
\cr
f(\beta) b^{\beta -2} &$\ell_1\ll b\ll\ell_0$ \cr
\cr
(\beta -2)^{-1}q_0^{2-\beta} & $b\gg\ell_0$\cr
}  
\ee
where we have considered the three limiting regimes 
$b\ll \ell_1$, $b\gg \ell_1$,
and $b\gg\ell_0$, and where
\be
g(\beta) = {{\Gamma (2-\beta/2)} \over {2^{\beta-1}}}  
\ee
\be
f(\beta) ={{\Gamma (2-\beta/2)\,\,\Gamma (\beta/2 -1)} \over {[\Gamma
(\beta/2)]^2\,\,\,2^{\beta-1}}}.
\ee
In this Appendix we are interested in using diffraction phenomena to estimate
the scattering measure.  Observations suggest that the relevant spectrum for the
interstellar medium has an outer scale $\ell_0 \equiv 2\pi/q_0$ that is 
larger than any diffraction scale or baseline that we consider.  

For spherical waves from a source embedded in the scattering medium we have
\be
D_s (b) = \int^D_0 ds\,\, 
\left [ \frac {\partial D_p (b/D)} {\partial s}\right ]_{b=sb/D}
\ee
\noindent
which implies, for a medium with $C^2_n =$ constant, that

\be
D_s (b) = D_p (b) \times \cases{ 3^{-1}\qquad\qquad b\ll\ell_1\cr
\cr
(\beta-1)^{-1}\quad\,\,b\gg\ell_1.\cr} 
\ee
If $b_e$ is the $1/e$ point of the visibility function
\be
\Gamma (b) = \exp [-D_{p,s} (b) /2], 
\ee
then the scattering diameter (FWHM) is
\be
\FWHM = {{2\sqrt{\ln 2}} \over {\pi}} \,\,\biggr( {{c} \over {\nu
\,\,b_e}}\biggr) 
\label{eq:FWHM}
\ee
from which it follows that the scattering measure for the plane wave case is

\be
SM_p = \cases{C_{\beta} \nu^{\beta}\,\,\FWHM^{\beta-2} \quad\quad\FWHM
\lesssim\theta_{\rm cross}\cr
\cr
D_{\beta} \nu^4\,\,\FWHM^2\,\,\ell_1^{\,4-\beta} \,\, \FWHM \gtrsim
\theta_{\rm cross}\cr} 
\ee
where


\begin{eqnarray}
C_{\beta} &=& \biggr( {{\pi} \over {2\sqrt{\ln 2}}}\biggr)^{\beta-
2}\,\,\big[ 4\pi^2\,\,r^2_e\,\,c^{\beta}\,f(\beta)\big]^{-1}\nonumber\\
D_{\beta} &=& \biggr( {{\pi} \over {2\sqrt{\ln 2}}}\biggr)^2\,\,\big[
4\pi^2\,\,r^2_e\,\,c^4\,\,g(\beta)\big]^{-1}.\nonumber
\end{eqnarray}
The angle $\theta_{\rm cross}$ is the boundary between the two regimes and is

\be
\theta_{\rm cross} = \biggr[ {{g(\beta)} \over {f(\beta)}}\biggr]^{1/(4-
\beta)}\,\theta_1  
\ee
where

\be
\theta_1 \equiv {{2\sqrt{\ln 2}} \over {\pi}}\,\,\biggr({{c} \over {\nu\,\,
\ell_1}}\biggr) 
\ee
is the scattering diameter if the $1/e$ half width of the visibility function
happens to be equal to the inner scale $\ell_1$ (cf.\ Eqn. \ref{eq:FWHM}).

For spherical waves the estimators for the structure function become

\be
SM_s = \cases{(\beta-1)\,\,C_{\beta}\,\,\nu^{\beta}\,\,\FWHM^{\beta-2},
\,\,\,\theta \lesssim \theta_{\rm cross}\cr
\cr
3\,\, D_{\beta}\,\,\nu^4\,\,\FWHM^2\,\,\ell_1^{4-\beta},\quad\quad\theta
\gtrsim\theta_{\rm cross}.\cr}  
\ee

\subsection{A2. Temporal Broadening}

Williamson (1972, 1975) has argued that temporal broadening of an impulse 
may be treated with stochastic ray tracing through extended random media even
though broadening is really a diffraction phenomenon.  Williamson demonstrated
the equivalence of ray tracing and a proper treatment of diffraction for a
statistically homogeneous medium having a Gaussian wavenumber spectrum for the
irregularities.  We assume the equivalence also holds for media with power law
wavenumber spectra.  Lambert \& Rickett (1999) have demonstrated this
equivalence using numerical integrations of Kolmogorov media.
Therefore, we relate the mean square scattering angle to
the temporal broadening time and  consider the (nontrivial) relationship of the 
mean square angle to the measureable angular diameter of the 
scattered brightness distribution. 

Following the discussion in CL91, Cordes \& Rickett (1998, hereafter CR98), 
and Lambert \& Rickett (1999),
we write the scintillation bandwdith $\dnud$ as
\be
\dnud = {C_1 \over 2\pi \langle \tau(D) \rangle}
\ee
with $C_1 = 1.16$  for a Kolmogorov wavenumber spectrum with
$\beta = 5/3$ and for a medium that uniformly fills the line of sight.   
 



The observable angular diameter corresponding to $\tau_d$ is
\be
\FWHM = \biggr( {{16 \ln 2 c \tau_d} \over {D}} \biggr)^{1/2} = 2.14\,\,{\rm
arc\,\,sec} \biggr( {{\tau_d} \over{D_{kpc}}}\biggr)^{1/2}  
\ee
which may be substituted into eqn (A14) to obtain the scattering measure.  For
measurements of the scintillation bandwidth, $\dnud$, eqn's (A21) and (A26)
may be used to solve for the scattering diameter,
\be
\FWHM 	= \left( 
		\frac {8\ln 2 c C_1} {\pi \dnud D}
	  \right)^{1/2} 
	= 0.919 \,\,{\rm mas} \left ({\dnud}_{\rm MHz} D_{kpc}\right)^{-1/2}. 
\ee
We now restrict ourselves to $\beta= 11/3$ for an inner scale 
$\ell_1 = 100\ell_{100}$km, for which  $f(\beta) = 1.118$ and
$g(\beta) = 0.877$.

The scattering measure estimators become for plane waves (extragalactic sources)

\be
SM_p = \cases{\biggr(\displaystyle{{{\FWHM} \over {0.128\,\,{\rm
arc\,\,sec}}}}\biggr)^{5/3}\,\,\nu^{11/3}_{GHz}\quad\quad\,\,\FWHM \lesssim
\theta_{\rm cross}\cr
\cr
\biggr(\displaystyle{{{\FWHM}\over{0.133\,\,{\rm arc\,\,sec}}}}\biggr)^2\,\,
\nu^4_{GHz}\,\,\ell_{100}^{1/3}\quad\FWHM \gtrsim\theta_{\rm cross}\cr}
\ee
while for spherical waves (Galactic sources)

\be
SM_s = \cases{ \biggr( \displaystyle{{{\FWHM} \over {0.071\,\,{\rm
arc\,\,sec}}}}\biggr)^{5/3}\,\,\nu_{GHz}^{11/3}\qquad\,\,\,\FWHM \lesssim
\theta_{\rm cross}\cr
\cr
\biggr( \displaystyle{{{\FWHM} \over {0.076\,\,{\rm arc\,\,sec}}}}\biggr)^2\,\,
\nu_{GHz}^4\,\,\ell_{100}^{1/3}\quad\,\,\FWHM \gtrsim\theta_{\rm cross}\cr}. 
\ee
The cross-over angle is
\be
\theta_{\rm cross} = 0.16\,\,
	{\rm arc\,\,sec}\,\,(\nu_{GHz}\,\,\ell_{100})^{-1},
\ee
corresponding to a temporal broadening time 
\be
\tau_{\rm cross} = 5.46 \, {\rm ms}\,\,D_{\rm kpc}\,\,(\nu_{\rm GHz}\,\,\ell_{100})^{-2} 
\ee
and to a  scintillation bandwidth
\be
\Delta\nu_{\rm cross} = 33.6\,{\rm Hz} \,\,D_{\rm kpc}^{-1}\,\,(\nu_{\rm GHz}\,\,\ell_{100})^{2}.
\ee

\end{document}

%% file: tab.neclumpN.P1.txt
${\rm 0540+23}$  &  $ -175.64$  &  $  -3.32$  &  $   0.63$  &  $   4.08$ \quad\quad  &  $\cdots$ \quad\quad\quad  &  $ 0.01$  \\ 
${\rm 0611+22}$  &  $ -171.21$  &  $   2.40$  &  $   0.65$  &  $  23.22$ \quad\quad  &  $\cdots$ \quad\quad\quad  &  $ 0.01$  \\ 
${\rm 0626+24}$  &  $ -171.18$  &  $   6.22$  &  $   0.65$  &  $   7.27$ \quad\quad  &  $\cdots$ \quad\quad\quad  &  $ 0.01$  \\ 
${\rm 0820+02}$  &  $ -138.01$  &  $  21.25$  &  $   0.55$  &  $   2.48$ \quad\quad  &  $\cdots$ \quad\quad\quad  &  $ 0.01$  \\ 
${\rm 0134-2937}$  &  $ -129.78$  &  $ -80.25$  &  $   0.53$  &  $   9.04$ \quad\quad  &  $\cdots$ \quad\quad\quad  &  $ 0.01$  \\ 
${\rm 0628-28}$  &  $ -123.05$  &  $ -16.76$  &  $   0.91$  &  $   9.39$ \quad\quad  &  $\cdots$ \quad\quad\quad  &  $ 0.01$  \\ 
${\rm GumI}$  &  $ -100.00$  &  $  -1.00$  &  $   0.50$  &  $ 110.51$ \quad\quad  &  $\cdots$ \quad\quad\quad  &  $ 0.14$  \\ 
${\rm GumII}$  &  $ -97.20$  &  $   2.70$  &  $   0.50$  &  $  28.36$ \quad\quad  &  $-0.59$ \quad\quad\quad  &  $ 0.03$ \\

%% file: iss_table_psr.tex
\begin{deluxetable}{lrrrrrrcrrr}
\tabletypesize{\footnotesize}
\tablecolumns{11}
\tablewidth{0pc}
\tablecaption{Scattering Values for Pulsar Lines of Sight \label{tab:psr.smtable}}
\tablehead{
\\
\multicolumn{2}{c}{Name}
\\
\\ 
\cline{1-2}  &&
\multicolumn{1}{c}{$\ell$} &
\multicolumn{1}{c}{$b$} &
\multicolumn{1}{c}{DM} &
\multicolumn{1}{c}{$\rm\hat D^{\tablenotemark{a}}$} &  
\multicolumn{1}{c}{$\nu$} &  
\multicolumn{1}{c}{Observable$^{\tablenotemark{b}}$} &  
\multicolumn{1}{c}{$\log$ Value$^{\tablenotemark{c}}$} &  
\multicolumn{1}{c}{$\log$ SM$^{\tablenotemark{d}}$} &
\multicolumn{1}{c}{Ref} 
\\
\multicolumn{1}{c}{ J2000} &
\multicolumn{1}{c}{ B1950} & 
\multicolumn{1}{c}{ (deg)} &
\multicolumn{1}{c}{(deg)} &
\multicolumn{1}{c}{(pc cm$^{-3}$)} &
\multicolumn{1}{c}{(kpc)} &
\multicolumn{1}{c}{(GHz)} &  
&
\multicolumn{1}{c}{(MHz or ms)} &
\multicolumn{1}{c}{(kpc m$^{-20/3})$} 
\\
}
\input psr.smtable.txt   
\\
\tablenotetext{a}{Distance estimates are based on NE2001.}
\tablenotetext{b}{Observables are the scintillation bandwidth, $\dnud$,
	defined as the half-width-at-half maximum of the intensity
	correlation function of the dynamic spectrum (c.f. Cordes 1986)
	and the pulse broadening time, $\tau_d$, that is approximately the
	$e^{-1}$ width of the pulse broadening function.}
\tablenotetext{c}{Scintillation bandwidths, $\dnud$, are in MHz while
pulse broadening times $\tau_d$ are in ms.  Values in reference 1
are scaled to 1 GHz using a Kolmogorov scaling law, 
$\dnud\propto\nu^{-22/5}$.  Errors are not quoted because they are
not reported uniformly in the different references and because
scintillation bandwidths in particular can vary markedly with
epoch (e.g. Bhat \etal\ 1999).} 
\tablenotetext{d}{Errors on $\log$ SM are not reported but would
receive contributions from errors in the observables and in the 
distance.  Typically, the error is $\Delta\log {\rm SM} \sim 0.3$.}
\tablenotetext{e}{Pulse broadening times reported in reference 12
are from an analysis of pulsars reported in reference 14.}
\end{deluxetable}

%% file: psr.smtable.txt
 \startdata
 \\
 $\rm 0528+2200$ & $\rm 0525+21 $ & $-176.14$ & $ -6.90$ &   50.9\quad\quad & 
            2.27 &  1.00 & $\Delta\nu_d$   & $ -0.070$ & 
          $ -3.39\quad\quad$ &  1 \\ 
 $\rm 0543+2329$ & $\rm 0540+23 $ & $-175.64$ & $ -3.32$ &   77.7\quad\quad & 
            3.54 &  1.00 & $\Delta\nu_d$   & $ -1.041$ & 
          $ -2.74\quad\quad$ &  1 \\ 
 $\rm 0534+2200$ & $\rm 0531+21 $ & $-175.44$ & $ -5.78$ &   56.8\quad\quad & 
            2.00 &  0.32 & $\tau_d$        & $ -0.638$ & 
          $ -2.64\quad\quad$ &  2 \\ 
 $\rm 0614+2229$ & $\rm 0611+22 $ & $-171.21$ & $  2.39$ &   96.7\quad\quad & 
            4.72 &  1.00 & $\Delta\nu_d$   & $ -1.022$ & 
          $ -2.86\quad\quad$ &  1 \\ 
 $\rm 0629+2415$ & $\rm 0626+24 $ & $-171.18$ & $  6.22$ &   84.2\quad\quad & 
            4.67 &  1.00 & $\Delta\nu_d$   & $ -1.092$ & 
          $ -2.80\quad\quad$ &  1 \\ 
 \\
 $\rm 0826+2637$ & $\rm 0823+26 $ & $-163.04$ & $ 31.74$ &   19.5\quad\quad & 
            0.37 &  1.00 & $\Delta\nu_d$   & $  0.980$ & 
          $ -3.60\quad\quad$ &  1 \\
\enddata
 \tablerefs{
 (1) Cordes (1986);
 (2) Cordes, Weisberg, \&  Boriakoff (1985);
 (3) Phillips \& Clegg (1992);
 (4) Johnston, Nicastro \& Koribalski (1998);
 (5) Ramachandran \etal\ (1997);
 (6) Mitra \& Ramachandran (2001);
 (7) Alurkar  \etal\ (1986);
 (8) Roberts \& Ables (1982);
 (9) Johnston (1990);
 (10) Costa, McCulloch, \& Hamilton (1991);
 (11) Gothoskar \& Gupta (2000);
 (12) Weisberg \etal\ (1990);
 (13) Foster (1990);
 (14) Clifton \etal\ (1992);
 (15) Dewey, Cordes, \&  Wolszczan (1988);
 (16) Cordes \etal\ (1990); Rawley (1986);
 (17) Camilo \& Nice (1995);
 (18) Fruchter \etal\ (1988);
 (19) Bhat, Rao \& Gupta (1999);
 (20) Nicastro \etal\ (2001) 
 }

%% file: iss_table_gal.tex
\begin{deluxetable}{lrrrrrrrrrr}
\tabletypesize{\footnotesize}
\tablecolumns{11}
\tablewidth{0pc}
\tablecaption{Scattering Values for Galactic Lines of Sight (Non-pulsars) \label{tab:gal.smtable}}
\tablehead{
\\
\multicolumn{1}{c}{Name} &
\multicolumn{1}{c}{$\ell$} &
\multicolumn{1}{c}{$b$} &
\multicolumn{1}{c}{$\rm\hat D$} &  
\multicolumn{1}{c}{$\nu$} &  
\multicolumn{1}{r}{$\theta_d$} &  
\multicolumn{1}{c}{$\log$ SM} &
\multicolumn{1}{c}{Ref} 
\\
&
\multicolumn{1}{c}{ (deg)} &
\multicolumn{1}{c}{(deg)} &
\multicolumn{1}{c}{(kpc)} &
\multicolumn{1}{c}{(GHz)} &  
\multicolumn{1}{r}{(mas)} &  
\multicolumn{1}{c}{(kpc m$^{-20/3})$} 
\\
}
\input gal.smtable.txt   
\\
\end{deluxetable}

%% file: gal.smtable.txt
 \startdata
 \\
 $\rm MonR2    $ & $-146.30$ & $-12.60$ &   0.80 &  1.66 & $      3.4$ & 
          $ -0.96\quad\quad$ &  1 \\ 
 $\rm NGC6334 H20$ & $  -8.74$ & $  0.67$ &   1.70 & 22.00 & $<     0.5$ & 
          $< 1.76\quad\quad$ &  2 \\ 
 $\rm NGC6334N OH1665(-12.6)$ & $  -8.58$ & $  0.65$ &   1.70 &  1.66 & $     18.0$ & 
          $  0.25\quad\quad$ &  3 \\ 
 $\rm NGC6334N OH1665(-7.3)$ & $  -8.58$ & $  0.65$ &   1.70 &  1.66 & $     19.0$ & 
          $  0.28\quad\quad$ &  3 \\ 
 $\rm NGC6334N OH1665(-11.5)$ & $  -8.58$ & $  0.65$ &   1.70 &  1.66 & $     31.0$ & 
          $  0.64\quad\quad$ &  3 \\ 
 \\
 $\rm NGC6334N OH1665(-12.2)$ & $  -8.58$ & $  0.65$ &   1.70 &  1.66 & $     15.0$ & 
          $  0.11\quad\quad$ &  3 \\
\enddata
 \tablerefs{
 (1) Hansen \etal\ (1993);
 (2) Trotter \etal\ (1998);
 (3) Kent and Mutel (1982);
 (4) Fix, Mutel, Gaume \& Claussen (1982);
 (5) van Langevelde \etal\ (1992);
 (6) Frail \etal\ (1994);
 (7) Yusef-Zadeh \etal\ (1999);
 (8) Davies \etal\ (1976);
 (9) Zhao \etal\ (1991);
 (10) Lo \etal\ (1981);
 (11) van Langevelde \& Diamond (1991);
 (12) Wilkinson, Spencer \& Nelson (1988) 
 }

%% file: iss_table_xgal.tex
\begin{deluxetable}{lrrrrrrrrrr}
\tabletypesize{\footnotesize}
\tablecolumns{11}
\tablewidth{0pc}
\tablecaption{Scattering Values for Extragalactic Lines of Sight \label{tab:xgal.smtable}}
\tablehead{
\\
\multicolumn{1}{c}{Name} &
\multicolumn{1}{c}{$\ell$} &
\multicolumn{1}{c}{$b$} &
\multicolumn{1}{c}{$\nu$} &  
\multicolumn{1}{r}{$\theta_d$} &  
\multicolumn{1}{c}{$\log$ SM} &
\multicolumn{1}{c}{Ref} 
\\
&
\multicolumn{1}{c}{ (deg)} &
\multicolumn{1}{c}{(deg)} &
\multicolumn{1}{c}{(GHz)} &  
\multicolumn{1}{r}{(mas)} &  
\multicolumn{1}{c}{(kpc m$^{-20/3})$} & 
\\
}
\input xgal.smtable.txt   
\\
\end{deluxetable}

%% file: xgal.smtable.txt
 \startdata
 \\
 $\rm 87GB0600+2957 $ & $-178.86$ & $  3.95$ &  1.00 & $      1.5$ & 
          $ -3.22\quad\quad$ &  1 \\ 
 $\rm 87GB0558+2325 $ & $-173.40$ & $  0.32$ &  1.00 & $      3.7$ & 
          $ -2.57\quad\quad$ &  1 \\ 
 $\rm 0611+131 $ & $-163.00$ & $ -2.20$ &  0.41 & $<    40.0$ & 
          $<-2.26\quad\quad$ &  2 \\ 
 $\rm 0622+147 $ & $-163.00$ & $  1.10$ &  0.41 & $<   160.0$ & 
          $<-1.26\quad\quad$ &  2 \\ 
 $\rm 87GB0621+1219 $ & $-161.14$ & $ -0.41$ &  1.00 & $      2.8$ & 
          $ -2.77\quad\quad$ &  1 \\ 
 \\
 $\rm 87GB0622+1153 $ & $-160.57$ & $ -0.33$ &  1.00 & $<    18.0$ & 
          $<-1.42\quad\quad$ &  1 \\
\enddata
 \tablerefs{
 (1) LC98a;
 (2) Dennison \etal\ (1984);
 (3) Resch (1974);
 (4) Rodriguez, Canto \& Moran (1982);
 (5) Trotter \etal\ (1998);
 (6) LC98c;
 (7) Lazio \etal\ (1999);
 (8) Readhead \& Duffet-Smith (1975);
 (9) Frail \etal\ (1994);
 (10) Fey \etal\ (1991);
 (11) Spangler, Mutel, Benson \& Cordes (1986);
 (12) Fey (1989);
 (13) Spangler \& Cordes (1998);
 (14) Mutel (1999) private communication;
 (15) Mutel \& Hodges (1985);
 (16) Geldzahler \& Shaffer (1979) 
 }